\documentclass[aps,showpacs,nofootinbib]{revtex4-1}
\usepackage{amsmath}
\usepackage{amsfonts}
\usepackage{amssymb,amscd,amsbsy}
\usepackage{epsfig}
\usepackage{color}

\usepackage{mathtools}

\newcommand{\bea}{\begin{eqnarray}}
\newcommand{\eea}{\end{eqnarray}}

\newcommand{\tr}{\textcolor{black}}
\newcommand{\kt}{k_{\rm B}T}

\newcommand{\cref}{c^{(2)}_{\rm ref}}

\newcommand{\khop}{k_{\rm hop}}
\newcommand{\ktum}{k_{\rm tum}}
\newcommand{\kxy}{k_{\rm 1\leftrightarrow 2}}
\newcommand{\kd}{k_{\rm tum}^{\rm d}}
\newcommand{\ku}{k_{\rm tum}^{\rm u}}
\newcommand{\kins}[1]{k^{{\rm ins}}_{#1}}
\newcommand{\ains}{\alpha_{{\rm ins}}}
\newcommand{\unit}{\mathit{u}}
\newcommand{\Dsph}{D_{\rm sph}}
\newcommand{\Lsph}{L_{\rm sph}}
 \newcommand{\DtwoDlatt}{D_{\rm 2D}^{\rm latt}}
 \newcommand{\DoneDlatt}{D_{\rm 1D}^{\rm latt}}
 \newcommand{\DtwoDcont}{D_{\rm 2D}^{\rm cont}}
\newcommand{\DthreeDcont}{D_{\rm 3D}^{\rm cont}}
\newcommand{\rhobulk}{\rho_{\rm 3D}}
\newcommand{\khopzero}{k_{\rm hop}^{\rm 0}}
\newcommand{\z}{$\hat z$}
\newcommand{\x}{$\hat x$}
\newcommand{\y}{$\hat y$}
\newcommand{\lbox}{l_{\rm Box}}

\newcommand{\rins}{r_{\rm ins}}
\newcommand{\rhoinc}{\rho^{\rm dep}}

\newcommand{\nicefrac}[2]{%
    \raise.5ex\hbox{#1}%
    \kern-.1em/\kern-.15em%
    \lower.25ex\hbox{#2}}

\begin{document}
\title{Monolayers of hard rods on planar substrates: II. Growth}

\author{M. Klopotek$^1$, H. Hansen-Goos$^2$, M. Dixit$^3$, T. Schilling$^3$, F. Schreiber$^{1}$ and M. Oettel$^{1}$}
\affiliation{ 
$^1$ Institut f\"ur Angewandte Physik, Eberhard Karls Universit\"at T\"ubingen, D--72076 T\"ubingen, Germany \\
$^2$ Institut f\"ur Theoretische Physik, Eberhard Karls Universit\"at T\"ubingen, D--72076 T\"ubingen, Germany \\
$^3$  Universit\'e du Luxembourg, Theory of Soft Condensed Matter, Physics and Materials Sciences Research Unit, L-1511 Luxembourg, Luxembourg
}


\begin{abstract}
{
Growth of hard--rod monolayers via deposition is studied in a lattice model using rods with discrete orientations and in a continuum model with hard spherocylinders. The lattice model is treated with kinetic Monte Carlo simulations and dynamic density functional theory while the continuum model is studied by dynamic Monte Carlo simulations equivalent to diffusive dynamics. The evolution of nematic order (excess of upright particles, ``standing--up'' transition) is an entropic effect and is mainly governed by the equilibrium solution, {rendering a continuous transition} (paper I, J.~Chem.~Phys.~{\bf 145},~074902~(2016)). Strong non--equilibrium effects (e.g. a noticeable dependence on the ratio of rates for translational and rotational moves) are found for attractive substrate potentials favoring lying rods. Results from the lattice and the continuum models agree qualitatively if the relevant characteristic times for diffusion, relaxation of nematic order and deposition are matched properly. Applicability of these monolayer results to multilayer growth is discussed for a continuum--model realization in three dimensions where spherocylinders are deposited continuously onto a substrate via diffusion. 
}
\end{abstract}

\maketitle
\section{Introduction}

The dynamic adsorption process of particles at surfaces or interfaces is interesting in the context of various fields in physics and chemistry, e.g.~(i) growth of thin metallic films (isotropic particles), (ii) formation of Langmuir 
monolayers~\cite{review-Kaganer-Mohwald-Dutta}, (iii) self--assembly of organic monolayers from solution or by {vapor phase}
deposition (anisotropic particles, mostly rod--like)~\cite{review-Schreiber,review-Schwartz} and (iv) growth of thin films of organic semiconductors by {vapor phase}
deposition (anisotropic particles)~\cite{review-OMBD-Schreiber,review-OMBD-Witte-Woell}. These examples have a strong motivation from applications in common (smooth coatings, functionalized surfaces, efficient organic solar cells), but also allow exploring the questions of structure formation away from equilibrium on a more fundamental level.

Theoretical research in field (i), growth of thin films with isotropic particles, has focused on a kinetic description in terms of 
an evolution of the time--dependent coverage and cluster size (island) distribution, entailing simple rules for particles adsorbing 
to or desorbing from islands, or the merging and break--up of islands \cite{krug_book}. 
A key tool to investigate and corroborate particular theoretical views has been the method of kinetic Monte Carlo (KMC) simulations which treats the time--evolution of a system through a stochastic sequence of individual, atomic events. It is rejection--free, i.e. one keeps track only of allowed events, which guarantees an efficient simulation of fairly large systems. As a result of numerous theoretical and simulation studies, a fairly detailed description of growth scenarios, island size distribution and island shape has become available, mainly in terms of scaling relations~\cite{krug_review,Ein13}.      

In the case of anisotropic particles it seems to be particularly important and worthwhile to study the interplay between the equilibrium phase diagram/equation 
of state and the dynamics of film formation. Already in 3D bulk rod--like particles exhibit numerous phases (liquid, nematic, smectic of various kinds, and crystalline)---a variety which may further increase when they are near a substrate. It is expected that 
{the structure of a film grown not too far from equilibrium also reflects the equilibrium phase diagram}. 
The classical model for molecular monolayers on an {\it unstructured} substrate are Langmuir layers [(ii) above], i.e. amphiphilic molecules on a liquid water surface. The typical finding is that of multiple structural phases characterized {\it inter alia} by different tilt angles~\cite{review-Kaganer-Mohwald-Dutta}. On solid surfaces, self--assembled monolayers (SAMs, (iii) above) are the 
prototypical system~\cite{review-Schreiber,review-Schwartz}. The substrate may be amorphous such as for the popular silanes on (oxidized) silicon, or crystalline such as for thiols on gold. The crystallinity of the substrate obviously introduces additional constraints and a potential having a periodic corrugation. The main structural phases which have been found are a ``lying--down'' ($\lambda$) phase and a ``standing--up'' ($\sigma$) phase, depending on the level of coverage. Importantly, the existence of these phases translates directly into growth behavior that is qualitatively different~\cite{PRB-Schreiber}. Specifically, (depending on growth conditions) the $\lambda$ phase appearing first with the $\sigma$ phase subsequently indicates a change in the kinetics of growth and gives rise to (at least) two regimes.

We note that the case of Langmuir layers (i.e. no underlying lattice) changes the situation, in that continuous lateral 
spacing {would be possible, in principle}, in contrast to e.g. SAMs of thiols on gold. This is one reason for differences in the phase diagram, but both have in common that multiple phases with different tilt structure are possible.

We also note that there are other important systems with angular degrees of freedom, namely, those related to organic molecular beam deposition (OMBD) of pentacene, diindenoperylene or other rod--like molecules employed in organic electronics ((iv) above)
\cite{Schr04,Witt04,Ruiz03,Ruiz04,Duerr03,Kow06}.

{In this context we suggest to analyze simplified models from the realm of soft matter science via} a theoretical and 
computational route which we believe to have potential for addressing the interplay of equilibrium phases and 
structure formation. 
Anisotropic particles are modeled by rods having simple, 
classical interactions on a discrete, cubic lattice. These may encompass steric exclusion (hard rods), mutual attractions and interactions with a substrate. 
{The equilibrium properties of such models (bulk or thin films) can be addressed by classical density functional theory 
and Monte--Carlo simulations, and serve as a reference for growth studies.}
Apart from the lattice system, rods are additionally modeled in continuous space. In a first instance we limit 
ourselves to steric exclusions and attractions with a substrate; we treat the formation of a monolayer of these rods on the substrate. 
This modeling approach implies drastic coarse--graining of both the particle--particle interactions as well as the orientations, which are restricted to solely three, {namely, one perpendicular and two parallel orientation with respect to a substrate}. Nevertheless, restricted--orientation models of hard rods already show a rich phase diagram~\cite{Mar04}, which  compares qualitatively well with that of unrestricted--orientation models~\cite{Esc08}. 

In a previous paper~\cite{Oet16}, we have investigated the equilibrium properties of lattice rods by classical density functionals 
from fundamental measure theory (FMT) and simulation. For the case of monolayers, a continuous $\lambda$--$\sigma$--transition 
{(``standing--up transition")}
has been found, which also persists in the case of finite substrate potentials. The agreement between FMT and simulation was found to be very 
good. We compared these findings to simulations of hard spherocylinders with continuous positional and orientational 
degrees of freedom and corresponding density functional theory (DFT) in the low--density limit. 
In this case, the continuous $\lambda$--$\sigma$--transition is found as well, but the scaling with rod aspect ratio is different 
from the lattice. {Nevertheless, there is good qualitative agreement between the lattice and continuum regarding
the degree of order in the monolayer as a function of density.}

{Dynamics can now be introduced by the assumption that the
growth of monolayers proceeds by a constant flux of particles onto the substrate}. Owing to the hard--core constraint, only rods that find an empty space on the substrate are adsorbed. Such a setup mimics the adsorption of rods from a reservoir (bulk solution or gas phase) at a higher chemical potential, or from a reservoir under the influence of a gravitational potential (providing constant flux). For treating such a monolayer growth scenario, we formulate a dynamic DFT model on the basis of FMT and employ KMC simulations. KMC growth--type simulations with anisotropic particles are much more complex than those with isotropic particles and have, therefore, found limited attention in the literature.
As in our previous work, we also employ Monte Carlo (MC) simulations of hard spherocylinders with continuous degrees of freedom; growth in this model is commeasured with that in the lattice model after matching the kinetic parameters.  

{Previous theoretical work on the deposition of anisotropic molecules can be found in  Refs.~\cite{Clancy06,Clancy07,Klepp15,Klepp-thesis,Klepp16,Muc11}. 
In the Clancy group, the specific examples of monolayer growth with pentacene, 1P and 2P molecules on different substrates were 
modeled with hard lattice 
dimers and trimers \cite{Clancy06,Clancy07} possessing sticky contact interactions. These were motivated by quantum chemical calculations. 
Emphasis was  put on exploring different growth patterns upon variation of temperature and substrate type, yet
the relation to equilibrium phases was not investigated. Kleppman et al. \cite{Klepp15,Klepp-thesis,Klepp16} investigate a mixed lattice--continuum
model for 6P on patterned substrates, exploring the feasibility to reproduce experimental findings with certain
simplified interactions. Toward the fine--end of the resolution scale is an all--atom study of pentacene growth on C60 \cite{Muc11}.
Keeping the atomistic details comes at the price of a limited particle number (on the order of 100). Evidence for a rather sharp
$\lambda$--$\sigma$ transition has been found. }

The structure of the paper is as follows: In Sec. \ref{sec:dft} on density functional theory, we recapitulate the lattice version of FMT for hard--rod mixtures and derive the dynamic DFT equations. Sec.~\ref{sec:kmc} introduces KMC simulations for anisotropic particles, where more specific details on the implementation used here are described in App.~\ref{app:kmc}. Sec.~\ref{sec:sphero} describes the simulations in the continuum model with hard spherocylinders. Results from the lattice and the continuum models for monolayer growth are presented in Sec. \ref{sec:results},
 and Sec.~\ref{sec:summary} gives a summary with discussion on possible experimental relevance as well as an outlook for future research.

\section{Density Functional Theory}
\label{sec:dft}

\subsection{FMT for lattice models}
  \begin{figure}
   \epsfig{file=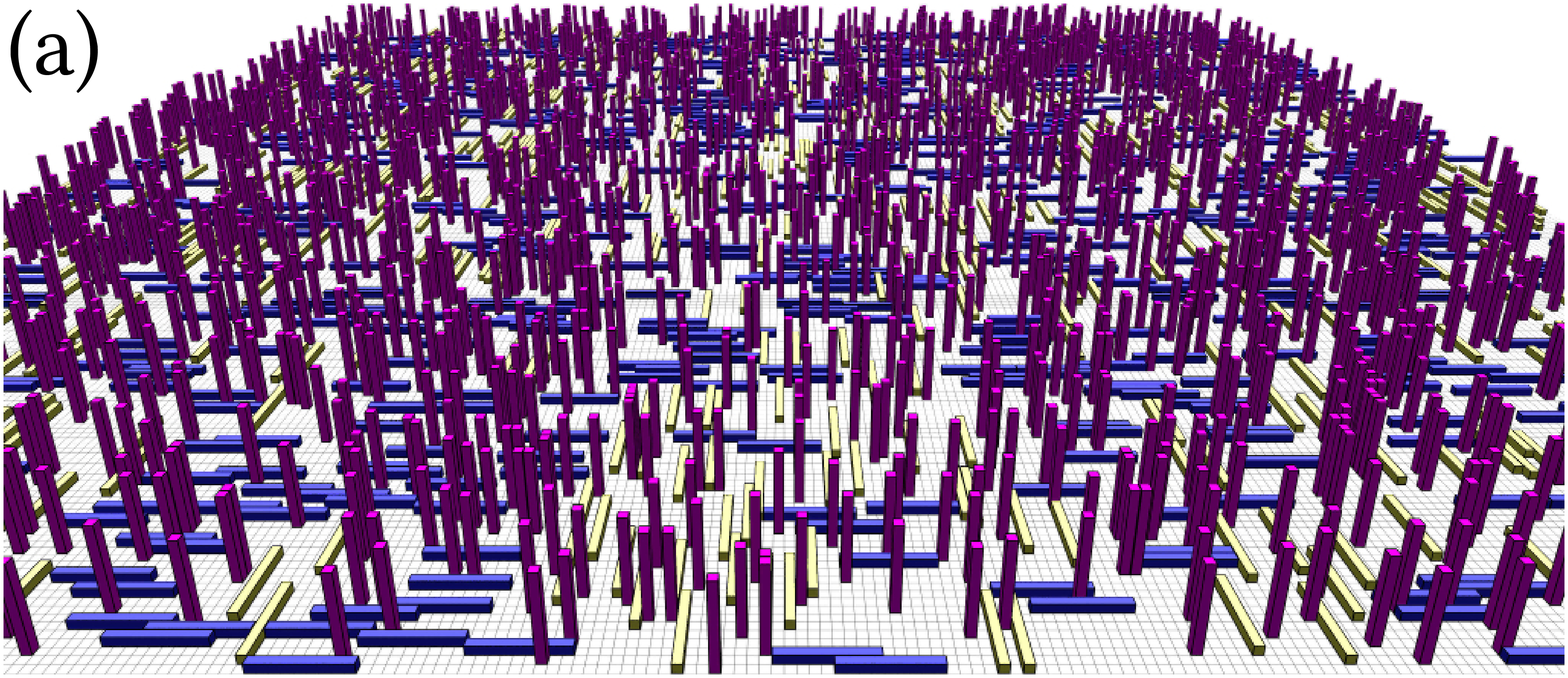,width=7.5cm} \hspace{2mm}
   \epsfig{file=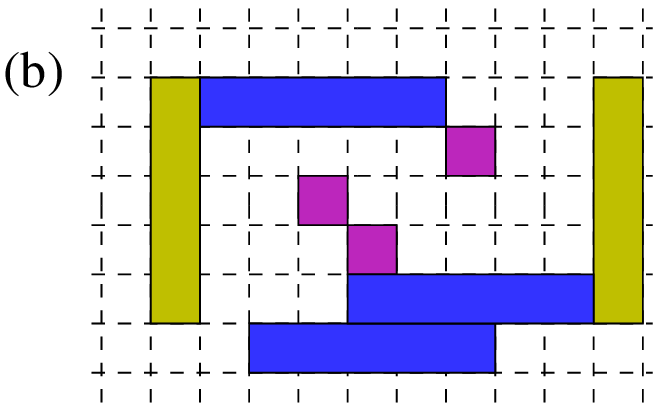,width=0.3\linewidth}
   \caption{(color online) Illustration of lattice model for hard rod monolayers, as seen in 3D (a) and projected on the $x$--$y$--plane (b). 
    Blue rods are oriented in $x$--direction, yellow rods in $y$--direction and
     magenta rods in $z$--direction.}
 \label{fig:latticesetup}
\end{figure}

A general FMT functional for hard rod mixtures on lattices with arbitrary dimensions has been derived by Lafuente and Cuesta~\cite{Laf02,Laf04}. In Ref.~\cite{Oet16}, we provide the basic definitions and examples for the functionals and their equilibrium properties for mono--component rods in two and three dimensions and in the monolayer case. 
In the present work, we only need the free energy functional for the homogeneous case for the monolayer. Rods with dimensions $1 \times 1 \times L$ (in lattice units) are confined to a substrate plane (square lattice) with their lower left corner (see Fig.~\ref{fig:latticesetup}(a)). Thus, the monolayer becomes a 2D ternary mixture of $1 \times L$ rectangles with two possible orientations in the substrate plane and $1 \times 1$ squares representing the upright rods (Fig.~\ref{fig:latticesetup}(b)).  The bulk number densities per unit square on the lattice are denoted by $\rho_1$, $\rho_2$ ($1 \times L$ rods with orientation in $x$-- and $y$--direction, respectively), and $\rho_3$ ($1 \times 1$ rods). The total density is $\rho=\rho_1+\rho_2+\rho_3$. The free energy density is given as a sum of an ideal gas part, excess part and external part:
\bea
    f &=& f^{\rm id} + f^{\rm ex} + f^{\rm ext}  \qquad \mbox{with}\label{eq_fdef} \\
    \beta f^{\rm id} & =& \sum_{i=1}^3 \rho_i \ln\rho_i - \rho \;,\label{eq_fiddef} \\
     \beta f^{\rm ex} &=& \Phi^{0d}(L(\rho_1+\rho_2)+\rho_3) -\Phi^{0d} \left( (L-1)\rho_1\right) -
                                         \Phi^{0d} \left( (L-1)\rho_2  \right) \label{eq_fexdef} \\
    \beta f^{\rm ext} &=& \sum_{i=1}^3 \rho_i V_i^{\rm ext} \;.
\eea 
Here, $\beta =1 /(\kt)$ is the inverse temperature {which will be set to 1 from now on}, and 
\bea
    \Phi^{0d}(\eta) &=& \eta + (1-\eta) \ln(1-\eta) \;.
\eea 
is the excess free energy of a zero--dimensional cavity (which can hold no or only one particle) depending on its average occupation $\eta \in [0,1]$. The substrate potential is specified  by the three constants $V_i^{\rm ext}$ which can be different from each other, in general.  

To characterize the behavior of the system, we introduce the order parameters
\bea
 Q & =& \frac{\rho_3-\frac{\rho_1+\rho_2}{2}}{\rho_1+\rho_2+\rho_3} \;, \nonumber \\
 S & = & \frac{\rho_1-\rho_2}{\rho_1+\rho_2}\;. \label{eq:orderparam3D}
\eea
$Q \not = 0$ signifies an excess ($Q>0$) or depletion ($Q<0$) of particles in the $z$--direction (nematic state) while $S \neq 0$ signals order in the $x$--$y$--plane orthogonal to the nematic director (biaxial state). Finite substrate potentials (with $V_1^{\rm ext}=V_2^{\rm ext} \not = V_3^{\rm ext}$) may introduce a nematic order $Q_{\rm id}$ for the very low--density ideal gas state. In Ref.~\cite{Oet16} we have found that $\delta Q = Q - Q_{\rm id} \propto \rho$ for low $\rho$, i.e. there is always continuous nematic ordering with increasing density and finite slope. For vanishing substrate potential, $\delta Q \propto \rho L^2$ for very long rods, and there is a reentrant transition to a biaxial state. These findings for $V_i^{\rm ext}=0$ are similar to those in Ref.~\cite{Mar14}, 
which treats a hard--rod model in restricted orientations but continuous translational degrees of freedom within FMT. 
{The effects of shape biaxiality have been investigated in Ref.~\cite{Mar15}, and rod--disk mixtures accordingly in Ref.~\cite{Mar16}.} 
{For corresponding results with continuum models, see Refs.~\cite{Oet16,Var16}.}

\subsection{Dynamic DFT on a lattice}
\label{sec:ddft}

{\subsubsection{Setup}}

The goal of our dynamic lattice DFT is to provide an {equation for} the time evolution of the observables $\rho_1$, $\rho_2$, and $\rho_3$ 
(or, equivalently, $\rho$, $Q$, and $S$) in a system driven out of equilibrium by particle deposition at constant rates. 
We limit our description to {this} tractable set of observables,
i.e., a given configuration of the system specified by these three observables stands for a much larger set of different microstates of the non--equilibrium system. Thus we cannot expect to reproduce trajectories of the system exactly. However, it is conceivable to gradually improve the description by refining the set of observables, thereby allowing for better discrimination of non--equilibrium configurations~\cite{Han05}.

Within the framework provided by the observables $\rho_1$, $\rho_2$, and $\rho_3$ the following formally--exact dynamic equations are readily obtained:
\begin{equation}
\label{eq_dyngen}
\frac{\partial\rho_i}{\partial t} = \alpha_i^{\text{ins}} p_i^{\text{ins}} + \sum_{j\neq i} \alpha_{j\to i}\rho_j p_{j\to i} - \sum_{j\neq i} \alpha_{i\to j}\rho_i p_{i\to j} \, ,
\end{equation}
where $i=1,2,3$. The constants $\alpha_i^{\text{ins}}$ correspond to the deposition rates of the individual orientation into an 
empty system. The parameters $\alpha_{i\to j}$ characterize the particle mobilities, i.e., the probability for a single particle 
of orientation $i$ in an otherwise empty system to change its orientation {and become a} particle of orientation $j$ is given by $\alpha_{i\to j} dt$. The complexity of the dynamics is {contained} in the probabilities $p_i^{\text{ins}}$ and $p_{i\to j}$, which denote the probability that an attempted particle deposition or orientational transition, respectively, is successful in a non--dilute system evolving along a certain non--equilibrium trajectory. These probabilities depend on the history of the system and generally cannot be expressed as functions of the $\rho_i$.

{Expressions for the probabilities $p_i^{\text{ins}}$ and $p_{i\to j}$ can be obtained by employing equilibrium--like
approximations, applicable for a situation where the deposition rates $\alpha_i^{\text{ins}}$ are very small compared to particle mobilities $\alpha_{i\to j}$.}
Using the excess chemical potential $\mu_i^{\text{ex}}$ of orientation $i$, 
\begin{equation}
  \mu_i^{\text{ex}} = \frac{\partial f^{\text{ex}}}{\partial \rho_i}
\end{equation}
from Eq.~\eqref{eq_fexdef}, we {use the thermodynamic definition of an insertion probability},
\begin{equation}
 p_i^{\text{ins}} = e^{- \mu_i^{\text{ex}}} = 
    \begin{cases}
     \frac{(1 - \eta)^L}{(1 - (L-1) \rho_i)^{L-1}}  & \text{for } i = 1,2 \\
     1 - \eta & \text{for } i = 3 \, ,
    \end{cases} 
 \label{eq:expmuex}
\end{equation}
where $\eta =  L (\rho_1 + \rho_2) + \rho_3$ denotes the packing fraction of the system, equivalent to the surface fraction of the substrate covered by a monolayer.

For the calculation of $p_{i\to j}$ we need to specify exactly how the orientation of a rod is changed under the given dynamics. To this end, we first consider a model where a change in orientation from $i$ to $j$ is realized in two steps. First, a rod with orientation $i$ is removed from the system and, second, a rod with orientation $j$ is inserted into the system at a random lattice site. We refer to these somewhat unrealistic dynamics as UNCO, denoting that removal and insertion of a rod are spatially {\em uncorrelated}. The quasi--equilibrium limit of $p_{i\to j}$ under the UNCO dynamics is readily obtained as $p_{i\to j} =  p_j^{\text{ins}}$. It can easily be checked that with these probabilities Eq.~\eqref{eq_dyngen} yields an equilibrium state with $\rho_i \propto e^{-\mu_i^{\text{ex}}}$ for $t\to\infty$, provided that no particles are deposited, i.e., $\alpha_i^{\text{ins}} = 0$. These are precisely the equilibrium particle densities following a minimization of the free energy $f=f^{\text{id}}+f^{\text{ex}}$ using Eqs.~\eqref{eq_fiddef} and \eqref{eq_fexdef}, w.r.t.\ the $\rho_i$.

\begin{figure}[tbp]
  \epsfig{file=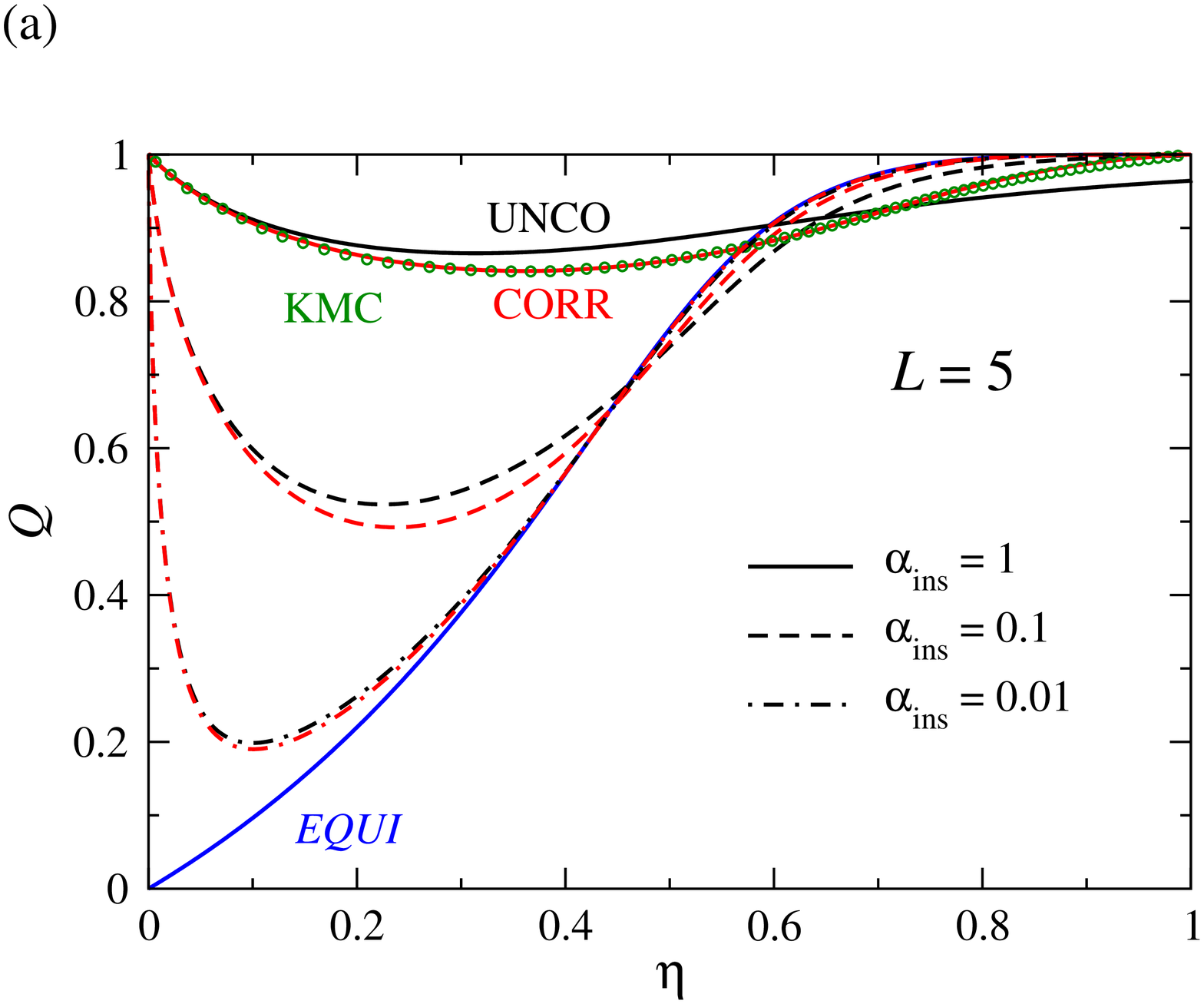, width=7.5cm} \hspace{3mm}
  \epsfig{file=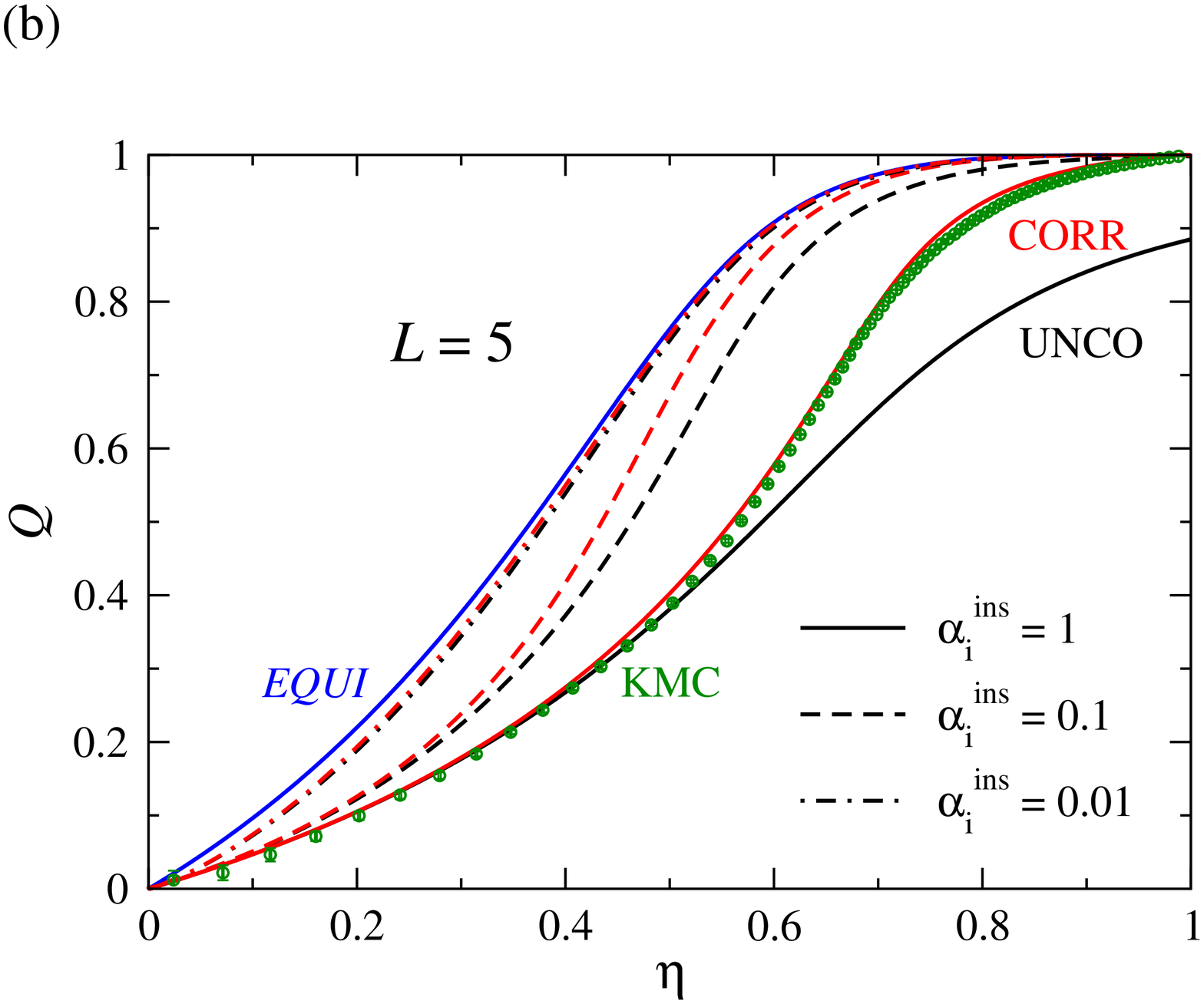, width=7.5cm} 
  \caption{ (color online) (a) Trajectories of the deposition of a monolayer of rods of length $L=5$ represented in the $(\eta,Q)$ plane, where $\eta$ denotes the covered surface fraction and $Q$ denotes the degree of nematic order in the monolayer. The system is initially empty ($\eta=0$) and rods are deposited with different rates $\alpha_{\text{ins}}$ measured relative to their rotational mobility. Rods are perpendicular to the substrate upon deposition. The blue curve corresponds to thermodynamic equilibrium. Results were obtained using the UNCO (black) and CORR (red) dynamic lattice DFT (see text). KMC simulations matched to the dynamics of $\ains=1$
{(fastest deposition)} were performed (green circles, {for a description see below}). 
Their error bars are smaller than the symbols. 
(b) Same analysis, but for deposition with random orientations $i=1..3$ inserted {with equal} rates $\alpha_i^{\text{ins}}$
(isotropic deposition). 
Error bars for KMC simulations (green) are displayed.} 
  \label{fig2}
\end{figure}

In order to compare the UNCO dynamic equations with our KMC simulations we make use of the fact that the equilibrium phase diagram obtained from Eqs.~\eqref{eq_fiddef} and \eqref{eq_fexdef} does not feature biaxiality for $L\leq 12$~\cite{Oet16}. 
We, therefore, make the assumption that $\rho_1=\rho_2$ also holds for the non--equilibrium setting of the rod--lengths 
studied in the simulations {(with $L = 5$ and 9 investigated below)}. We consider two different modes of particle deposition: (i) perpendicular deposition and (ii) isotropic deposition. The corresponding deposition rates are (i) $\alpha_1^{\text{ins}} =\alpha_2^{\text{ins}} = 0$, $\alpha_3^{\text{ins}} = \alpha_{\text{ins}}$ and (ii) $\alpha_1^{\text{ins}} = \alpha_2^{\text{ins}} = \alpha_3^{\text{ins}} =\frac{1}{3} \alpha_{\text{ins}}$. Time is measured relative to particle mobility, which we assume to be isotropic with $\alpha_{i\to j} = 1$, where $i\neq j$. Results of the UNCO model are obtained by solving the set of differential equations numerically for an initially empty system. Figure \ref{fig2} shows the UNCO trajectories of the system for different deposition rates $\alpha_{\text{ins}}$ in the $(\eta, Q)$ plane resulting from perpendicular and isotropic deposition. Since the behavior for different rod--lengths $L\le 12$ is found to be qualitatively the same in the model, we limit ourselves at this point to the case $L=5$. Results for rod--lengths $L = 9$  are shown in Sec.~\ref{sec:results}, where we compare the dynamic DFT results to simulations.

It is interesting to note that in the long--time limit $t\to\infty$ the UNCO dynamics do not necessarily generate a configuration in which all the rods stand up, i.e. for sufficiently fast deposition we find $Q<1$ while $\eta\to 1$. This is reflected in the UNCO dynamic equations being stationary for $\eta = 1$, irrespective of the value of $Q$, thereby allowing for a fully--covered surface with a certain fraction of rods still in the $\lambda$ orientation (i.e.\ lying down). While somewhat counter--intuitive, this behavior is rooted in the non--locality of the UNCO dynamics. Once a $\lambda$ rod is chosen for a change in orientation, these particular dynamics attempt to insert the rod after reorientation at a {\em random} site of the lattice. For sufficiently large $\eta$ this insertion is almost always impossible, even for a rod in the $\sigma$ orientation (i.e.\ standing up). As a result, the rod chosen to perform the move remains at its initial site in $\lambda$ orientation. {Consequently}, the system can remain locked 
{in a fully--packed configuration}, preventing it from switching out all the rods in the system to those with a perpendicular orientation (to the surface).

More realistic {\em local} dynamics is provided by the CORR model, which performs changes in orientation locally and takes {\em correlations} at the given site into account. It is based on the simple observation that if a transition from a $\lambda$ to $\sigma$ orientation is done locally, the move is always accepted since a rod lain down on the substrate automatically guarantees room for it to stand up at the same location. Hence, in the CORR model, we employ $p_{1\to 3} = p_{2\to 3} = 1$. In order to recover the correct equilibrium behavior in the stationary state without particle deposition, we must have $p_{3\to i} = e^{-\mu_i^{\text{ex}} + \mu_3^{\text{ex}}}$, where $i = 1, 2$. The remaining transition probabilities are the same as in the UNCO model. Assuming no biaxiality ($S=0$), results are obtained by solving the set of differential equations numerically for various depositions rates, considering both perpendicular and isotropic deposition. In Fig.~\ref{fig2} we show results of the CORR model for rods of length $L=5$. In particular, the theory predicts $Q=1$ in the limit $\eta = 1$, meaning that in the long--time limit with full surface coverage all the rods are in $\sigma$ orientation. It can easily be shown from the dynamic equations that, in contrast to the UNCO model, stationarity in the CORR model requires $\eta = 1$ {\em and} $\rho_1=\rho_2=0$. This implies $\rho_3 = 1$ and, therefore, $Q=1$.


Figure~\ref{fig2} includes data from KMC simulations (Sec.~\ref{sec:kmc}) {with matching dynamic parameters} 
(Sec.~\ref{sec:matchddft} below). We note that for the error bars, here and everywhere else, KMC data is first averaged into bins; thereafter the binned data is averaged over six independent runs. Exceptions are for $\alpha < 10^{-3}$, where data is collected from a single run; they are otherwise noted. The CORR model appears to give an excellent description of the dynamics of rods of lengths $L = 5$, particularly in the case of perpendicular deposition. Sec.~\ref{sec:results} compares the dynamic DFT results with our 
KMC simulations further for rod--length $L=9$.

\subsubsection{Finite substrate potentials}
\label{sec:ddft_vext}

{We consider} the case where the substrate interacts via an attractive potential of strength {$\epsilon$ per segment
touching the substrate}:
{
\bea
  V_i^{\rm ext} = \left\{ \begin{matrix} - \epsilon  & \qquad (i=3) \\
                                         -L\epsilon & \qquad (i=1,2) \end{matrix}      \right. 
\;.
\label{eq:vext}
\eea
}
Here, the rotational mobilities $\alpha_{i\to j}$ have to be partially modified. While the mobilities in the substrate 
plane remain unchanged (i.e.\ unity in the present normalization) the attractive interaction suppresses transition from a
$\lambda$ to $\sigma$ orientation, and within the present dynamics we have $\alpha_{i\to 3}=e^{-\epsilon(L-1)/2}$ for $i=1,2$. 
On the other hand, a transition from a $\sigma$ to $\lambda$ orientation is promoted, leading to modified mobilities $\alpha_{3\to i}=e^{\epsilon(L-1)/2}$ for $i=1,2$. In both the UNCO and the CORR models, these modified mobilities lead to stationary points for $\alpha_{\text{ins}}=0$, which are identical to the equilibrium properties obtained by minimization of the free energy functional with the 
appropriate external potential in Eq.~(\ref{eq:vext}). Note that we will study the scenario of an 
attractive substrate only in the case of perpendicular deposition, which means we may leave the insertion rate unmodified.

\subsubsection{Quasi--equilibrium growth}
\label{sec:quasieq}

When the flux rate is infinitely slow compared to all other kinetic parameters in the monolayer, every moment of 
growth is fully described by thermodynamic equilibrium. The change in density of species $j$ {through deposition} within 
time step $dt$ is proportional to the flux rate as
\bea
d\rhoinc_j = {\alpha^{\rm ins}_j} e^{-\mu_j^{\rm ex} }dt\;,
\label{eq:drhoinc}
\eea
{The deposited particles become redistributed instantaneously ($d\rhoinc_j \to d\rho_j$) with conservation
of the total number of particles},
\bea
d\rhoinc = \sum\limits_j d\rhoinc_j = \sum\limits_j d\rho_j = d\rho \;,
\label{eq:drho}
\eea
such that {the total chemical potential $\mu = \mu_j$, as well as the increments $d\mu = d\mu_j$,} are constant
{and equal among all species}. 
Here, $\mu_j = \ln \rho_j + \mu^{\rm ex}_j$. 
{We define
$r_{i j} = \frac{\partial \mu_i}{\partial \rho_j}$
and, thus,
\bea
d\mu = d\mu_i = \sum\limits_{j=1}^{3}r_{ij}d\rho_j \qquad (i=1..3)\;. 
\eea
}
{In our system $\rho_1 = \rho_2$.
Solving for the two independent density increments  we obtain
\bea
d\rho_i &=& A_i d\rho  \qquad (i=1,3)
\label{eq:drhoeq}
\eea
where $A_1 = \frac{r_{13} - r_{33}}{2 r_{13} + r_{31} +r_{32} - r_{11} - r_{12} - 2r_{33}}$, $A_3=1-2A_1$ and 
$d\rho$ is defined through Eqs.~(\ref{eq:drhoinc}) and (\ref{eq:drho})}.
{The total time increment in $d\rho$ can be re--scaled, $dt^\star=\ains dt$, such that the coupled system of equations in 
(\ref{eq:drhoeq}) does not depend on the total flux $\ains$ anymore. The solutions $\rho_i(t^\star)$ can then
be found through numerical integration.}

%

\section{Kinetic Monte Carlo Simulations}
\label{sec:kmc}

KMC is suited for simulating dynamical systems that can be characterized by {a finite number of} elementary processes occurring with different rates (denoted `events'). 
An underlying assumption is that each event $j$ having a rate $k_j$ occurs via a Poisson--process with mean waiting time 
$1/k_j$. As events occur independently, the total random process of waiting for \emph{any} among all events is also Poissonian 
with a mean waiting time $1/{\sum_j k_j}$~\cite{BKL,Ficht91,Levi97,Adam99,Jansen12}.
Specifically, this probability distribution of waiting times has the form $P^{\rm wait}(t) = e^{-{\sum_j k_j}/t}$. 

In each KMC iteration step, a single, currently--allowed event having rate $k_i$ is chosen randomly among all such events with a relative probability $\frac{k_i}{k_{\rm tot}(\lbrace \mathcal C\rbrace)}$ , $k_{\rm tot}(\lbrace \mathcal C\rbrace) = \sum_j k_j$, where $\left.\lbrace k_j \rbrace_j\right\vert_{\lbrace \mathcal C\rbrace}\ni k_j$ is the full list of allowed events at this configuration. ({Note that this list \emph{could} include \emph{forbidden} events~\cite{Adam99}, but at the computational cost of rejecting them.}) KMC is therefore effectively `rejection--free', at least in the variant of the algorithm used here, first proposed in~\cite{BKL}. The waiting time since the last event, i.e. the increment of time, is chosen according to the distribution $P^{\rm wait}(\Delta t,\lbrace \mathcal C\rbrace)$, employing $\Delta t = -1/k_{\rm tot}(\lbrace \mathcal C\rbrace) \ln u$, with $u \in (0,1]$ chosen randomly and uniformly. The chosen event is executed. The list of allowed events must be updated according to the new configuration by adding newly--allowed events and removing forbidden ones.

This tracking of allowed and forbidden events makes KMC non--trivial, {illustrated here for the case} of hard--core particles: 
An event might become forbidden, for example, if a nearest neighbor rod is blocking the hopping or tumbling move of a rod. Also, a new event must be added to the list once the nearest neighbor(s) in the way moves away from the rod. We implement a detection system that tracks proper neighborhood patterns. Such a system becomes increasingly complex the higher the degree of anisotropy of the particles. 
Our algorithmic approach (see App.~\ref{app:kmc}) can be extended to general hard--core lattice systems. 
{As one sees}, the rejection--free bonus of KMC comes at the cost of algorithmic complexity 
{to eliminate forbidden moves}.

The kinetics of our lattice model ({square lattice in the \x--\y--plane [substrate] with unit length $\unit$, size $M \times M = 256^2$, periodic boundary conditions}) 
is characterized by the rates of the allowed single--particle processes. The first rate is $\khop^0$ for an explicit hopping 
process of a rod of orientation $i$ ($i=1..3$) on the substrate, translating it by one lattice site in any of 4 directions. This process may occur regardless of the orientation, and the rates are identical. The second rate $\ktum$ is ascribed to a tumbling process, which changes the orientation of a rod. Here, the rod is assumed to rotate around one of its ends. Specifically, the tumbling process is split into two types---the first, a tumble `upward' into the \z--direction from a lying orientation ($i =1, 2$) to a standing one ($i=3$). This rate is denoted $\ku$. The second is a tumble `downward' into the \x--\y--plane from a standing orientation to a lying one. This is denoted $\kd$. The third rate $\kxy$ is the in--plane rotation between orientations 1 and 2 about the rod midpoints. This constrains our investigations to rod--lengths $L$ of odd number. All rates are in units of inverse time. 
The final rate (orientation--specific) is $\kins{i}$ for a random influx of rods of orientation $i$, in units of inverse time 
multiplied by $\unit^2$. This influx of rods {(corresponding to the insertion rate in the DDFT model)} is implemented as a 
random appearance of rods of orientation $i$ at constant rate $\kins{i}$ per lattice site, whereby the move is rejected 
if overlap occurs, {i.e. the rod ``disappears''}. There is hence a monotonic, but non--linear relationship 
between number density $\rho$ and simulated time (see Fig.~\ref{fig:rho_t}(a) below). 

\subsection{Matching to DDFT}
\label{sec:matchddft}

\begin{figure}
 \epsfig{file=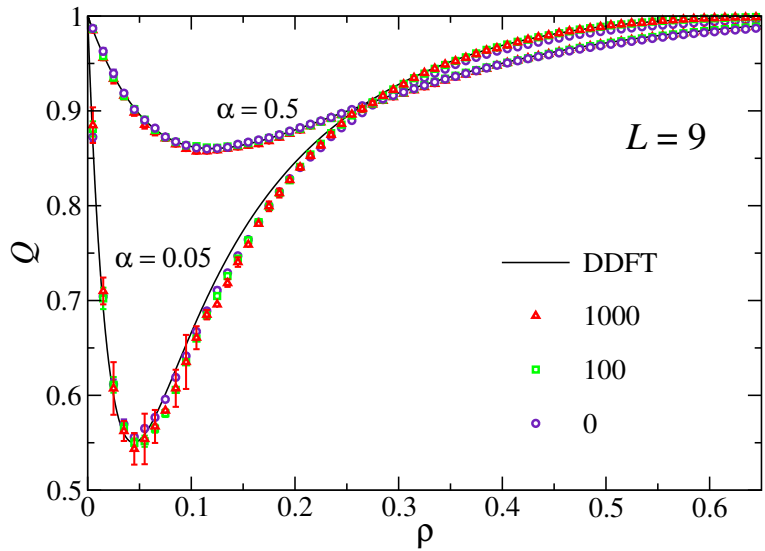,width=10cm}
\caption{(color online) Dependency of growth dynamics of the monolayer on the kinetic parameter $\khopzero$: 
There is virtually none. The total number density is $\rho$ and $Q$ denotes the degree of nematic order in the monolayer. 
Data with error bars is from KMC simulations with $L=9$, perpendicular deposition, and two cases of the growth parameter 
$\alpha=0.5,0.05$ (Eq.~(\ref{def:alpha})) and different values of $\khopzero/\ktum$. The solid curves are calculated by DDFT (CORR). The data set $\lbrace\khopzero/\ktum=1000,\alpha=0.05\rbrace$ is averaged over two independent runs instead of six. }
\label{fig:khop}
\end{figure}

In the following we only consider the case where $\ku=\kd\equiv\ktum$ {(no substrate potential)}. 
The rates for the tumbling and deposition process are related to the rates defined in the DDFT equation (\ref{eq_dyngen}) as follows:
\bea
    \alpha_{i \to j}  &\leftrightarrow& 2 \ktum \qquad\mbox{all combinations with $i \neq j$} \\
    \alpha_i^{\text{ins}} &\leftrightarrow& \kins{i} \;.
\eea
where the relation means equality up to the same constant factor. The first relation holds since we have fixed $\alpha_{i \to j} = 1$ in DDFT globally. The factor two arises from the fact that the rods can rotate into each orientation in one of two rotational directions. Since one of the rates can be used to define the time scale, a growth process only depends on ratios of rates. 
{As introduced before, we consider 
vertical deposition ($\alpha_i^{\text{ins}} = \ains\delta_{i,3}$) or isotropic deposition 
($\alpha_i^{\text{ins}} = \frac{1}{3} \ains$) with the total deposition rate $\ains$}; 
the same deposition rate holds for the KMC model via $k_{\rm ins} = \sum_i \kins{i}$. We also assume isotropic transition rates $\alpha_{i \to j}$ (see Sec.~\ref{sec:ddft}), analogous to $\ku=\kd=\ktum=\kxy$ in KMC. 
DDFT predicts that there is no dependence of our observables on $\khopzero$. This is indeed what we also observe in KMC 
(see Fig.~\ref{fig:khop}). 
Our matching condition is hence set by the single independent variable $\alpha$ characterizing the growth dynamics:
\bea
\label{def:alpha}
\alpha := \frac{k_{\rm ins}}{2 \ktum} = \frac{\sum_i\kins{i}}{2 \ktum}\equiv \frac{\sum_i \alpha_i^{\text{ins}}}{\alpha_{i \to j}} =\frac{\ains}{1}  \;.
\eea
{This variable is different from the single variable, commonly denoted $F/D$, characterizing growth with isotropic particles,
where $F$ is the incoming flux rate and $D$ is the diffusion constant in the substrate plane.
For our KMC model, the translational diffusion constant $D\equiv D_{\rm 2D}^{\rm latt}$ in the dilute limit 
{(monolayer density close to zero)} is determined by 
both $\khopzero$ and $\ktum$:
\bea
\label{eq:D_latt}
  \DtwoDlatt/{\unit}^2 &=&   \left( \frac{1}{2}\ku + \frac{1}{1+2\frac{\kd}{\ku}}\left( \kd -\frac{1}{2}\ku \right) \right)\frac{(L-1)^2}{4} + \khopzero\\
			&\equiv& 	\khop + \khopzero \,,
\label{eq:khop}
\eea
see App.~\ref{app:derivation_D2Dlatt}. One sees that hopping and tumbling (through an effective hopping rate $\khop$)  contribute to diffusion independently.}


\subsection{Results: DDFT vs. KMC}
\label{sec:ddft_kmc}

\subsubsection{The case $\ku = \kd$ (no substrate potential)}
\begin{figure}
 \epsfig{file=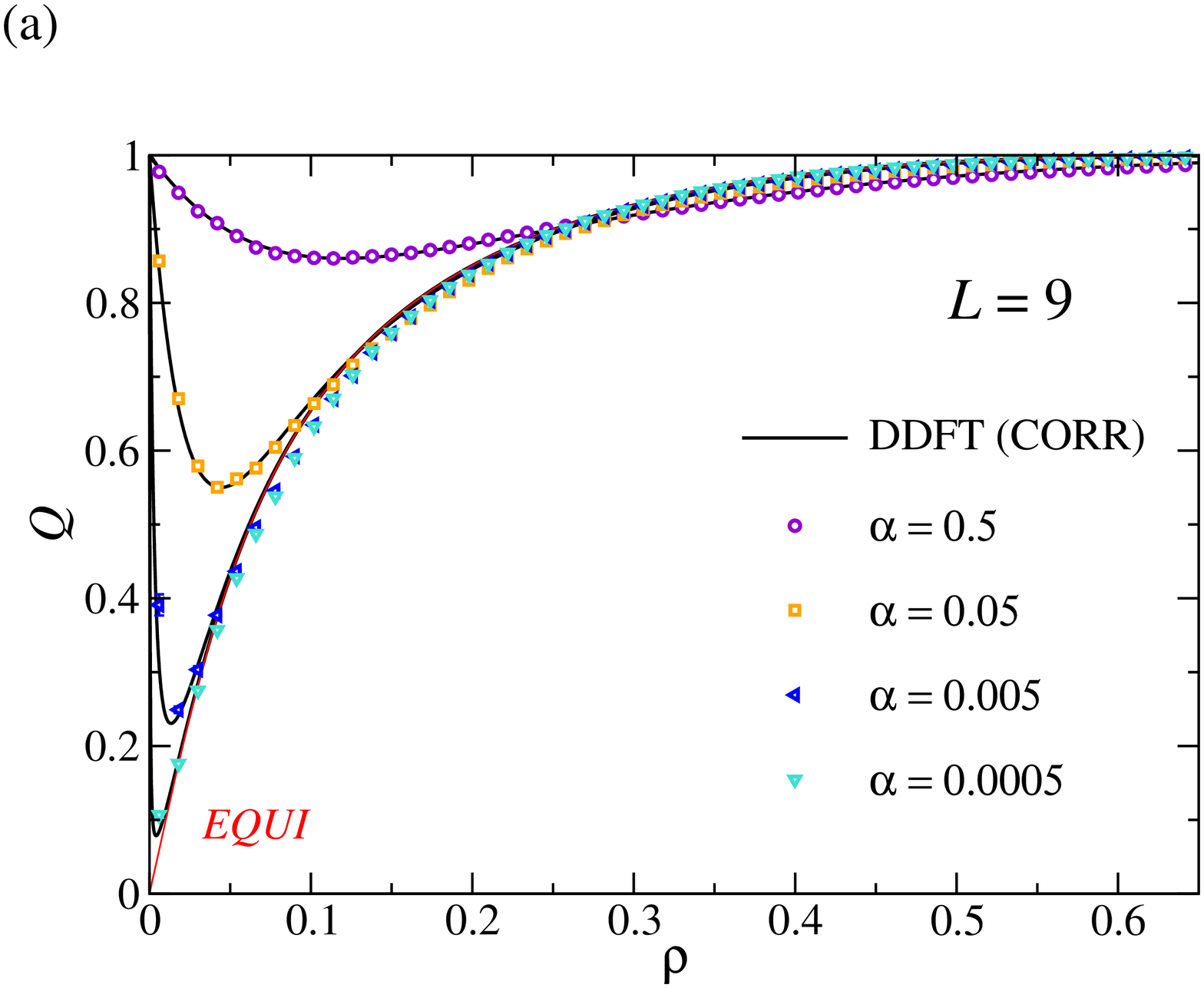,width=7.5cm}\hspace{3mm}
 \epsfig{file=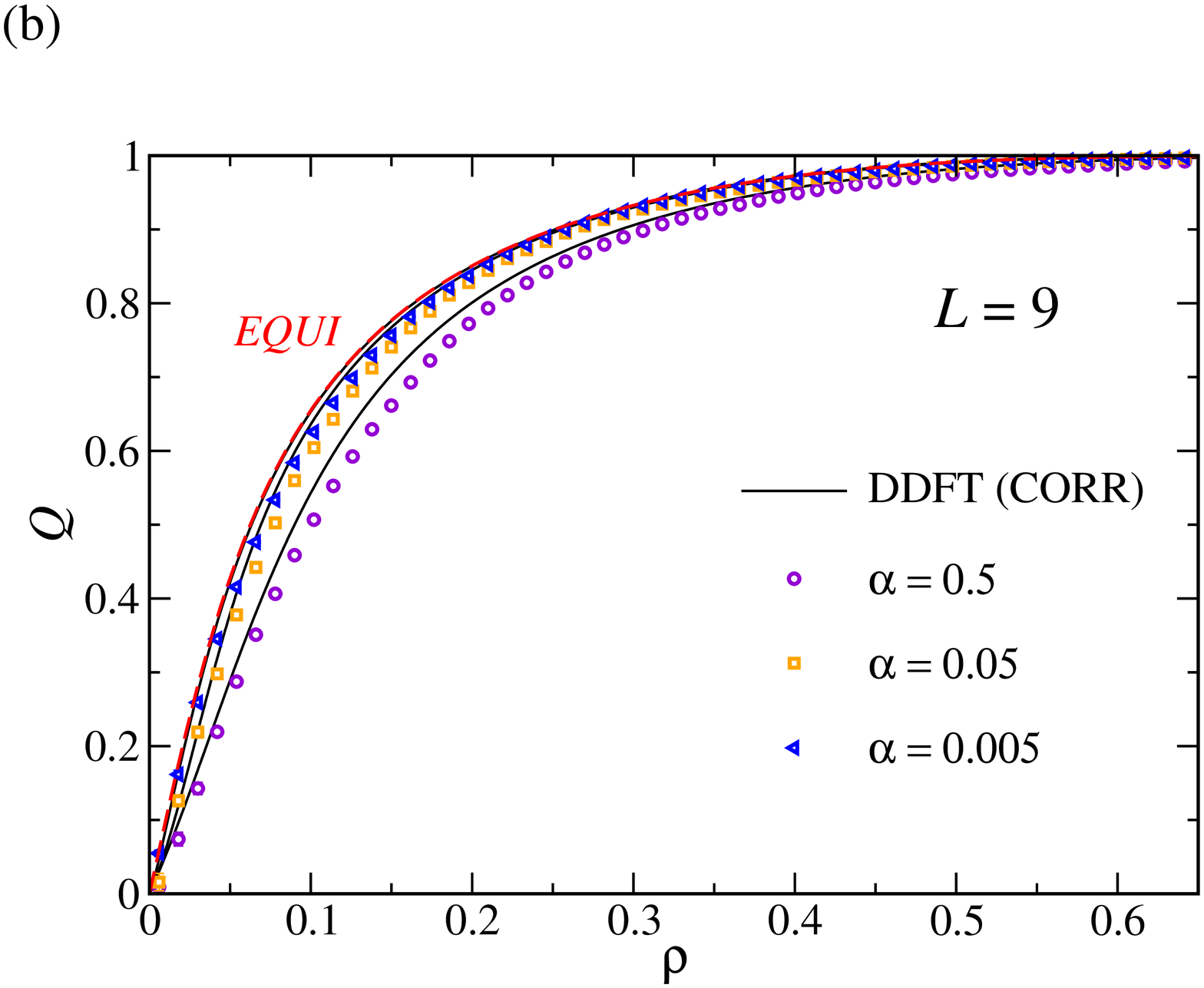,width=7.5cm}
\caption{(color online)  Trajectories of the deposition of a monolayer of rods of length $L=9$ represented in the ($\rho$, $Q$) plane,
for a varying growth parameter $\alpha$. 
Plotted are calculations with KMC (symbols with error bars) and DDFT (CORR) (black lines) with (a) perpendicular deposition and 
(b) isotropic deposition. The red curves (EQUI) correspond to solutions from  equilibrium DFT.}
\label{fig:ddft_kmc}
\end{figure}

\begin{figure}
 \epsfig{file=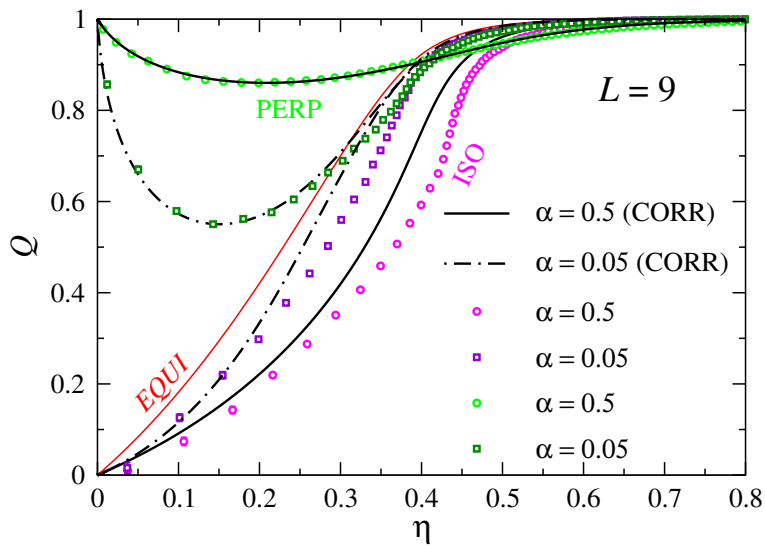,width=10cm}
\caption{(color online)  Trajectories of the deposition of a monolayer of rods of length $L=9$ represented represented in the ($\eta$, $Q$) 
plane for $\alpha=0.5, 0.05$. Curves are DDFT (CORR), symbols are KMC data; green (PERP) indicates perpendicular deposition, 
pink/violet (ISO) indicates isotropic deposition.  The red curve (EQUI) corresponds to  solutions from equilibrium DFT.}
\label{fig:ddft_kmc_eta}
\end{figure}

\begin{figure}
 \epsfig{file=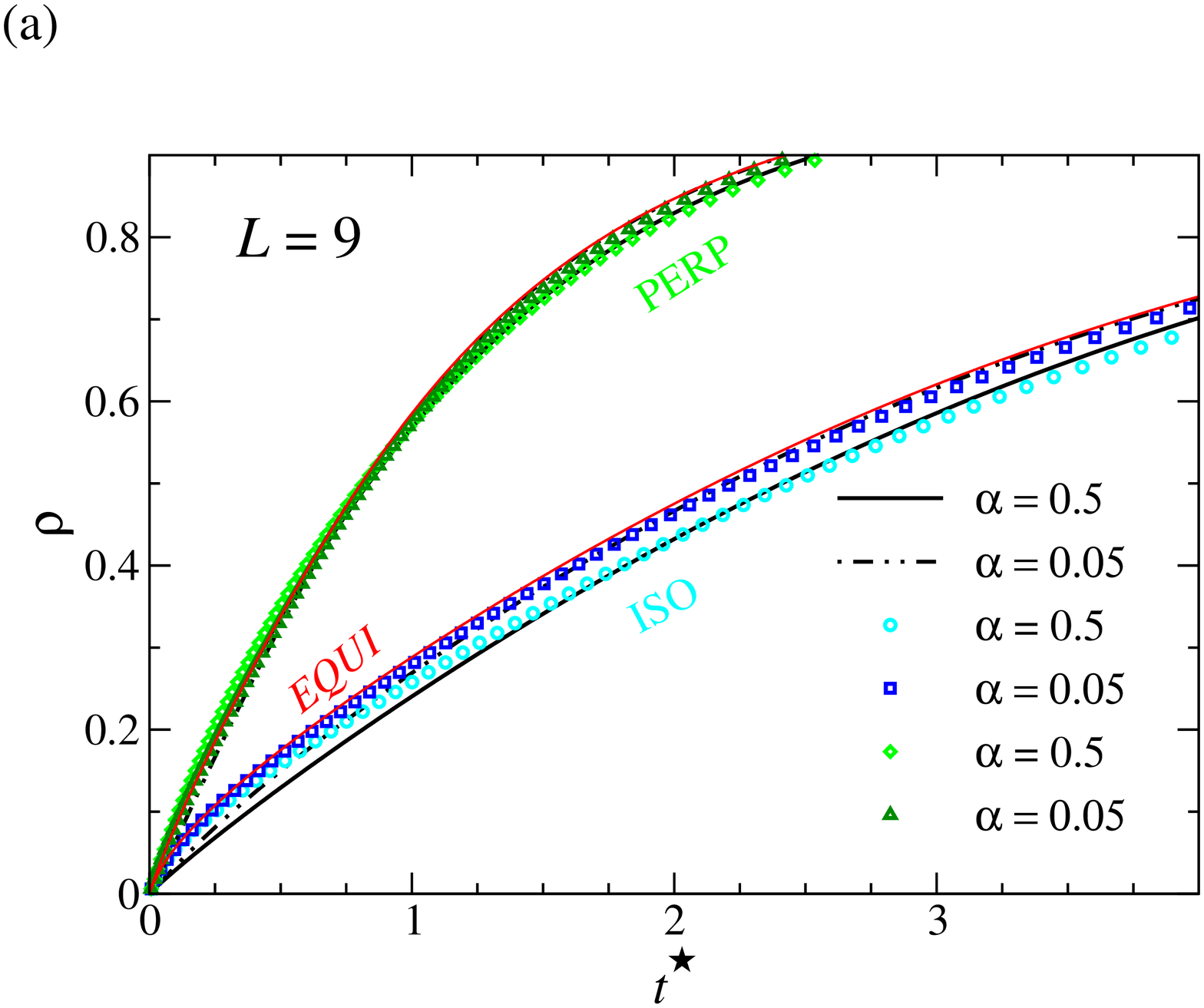,width=7.5cm}\hspace{3mm}
 \epsfig{file=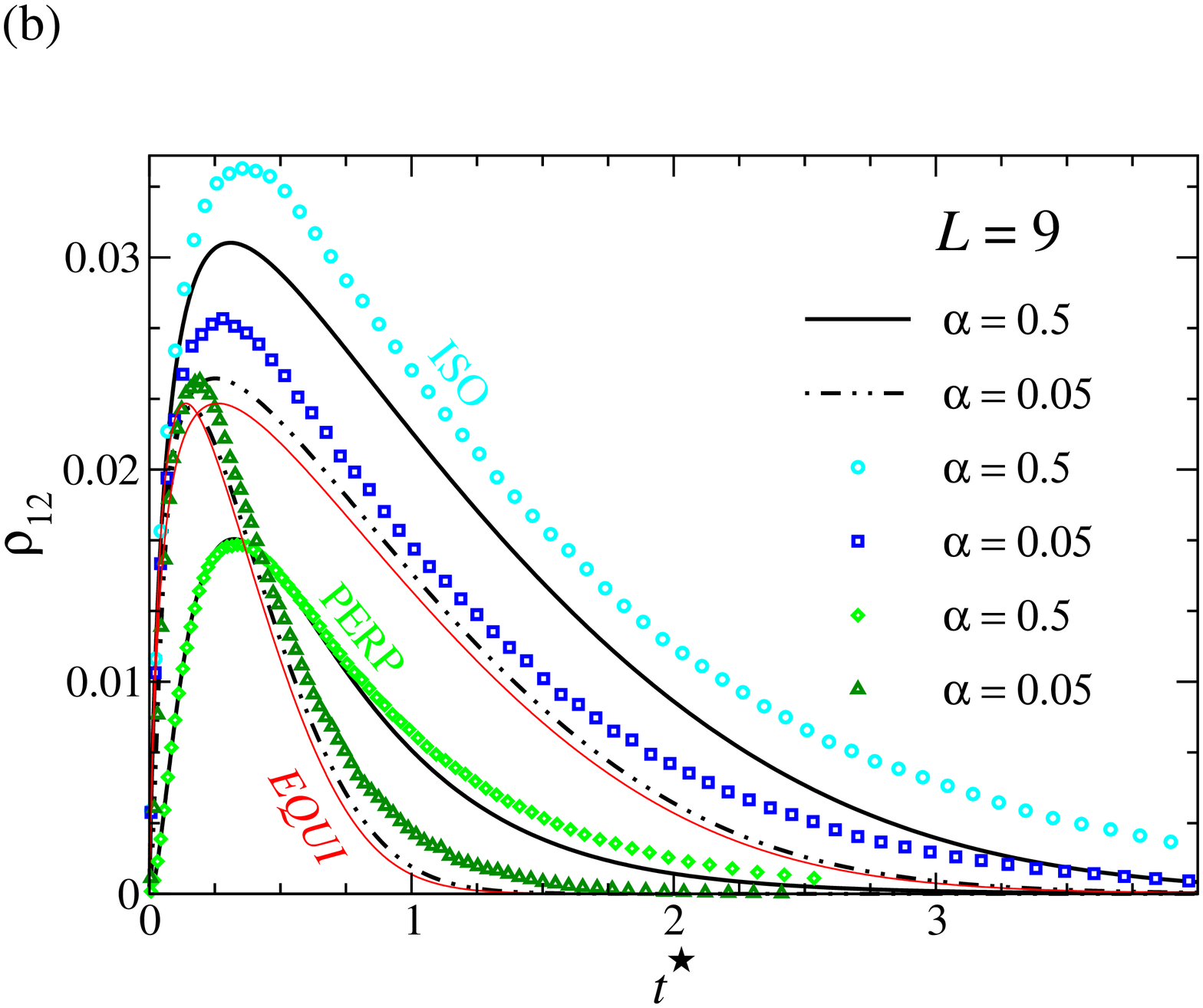,width=7.5cm}
\caption{(color online)  Evolution of number densities during growth in a monolayer of rods of length $L=9$ with growth parameters $\alpha=0.5,0.05$; perpendicular-- (PERP, green) as well as isotropic--deposition growth (ISO, cyan/blue) is calculated via KMC (symbols with error bars) and DDFT (CORR) (black lines). (a) Total number density $\rho$ and (b) number densities of lying rods $\rho_{1 2} = \rho_1 + \rho_2$ versus time re--scaled with the flux rate $t^\star = \alpha_{\rm ins} t$. The red curves (EQUI)---shown for both deposition types correspond to infinitely--slow, quasi--equilibrium growth calculated from DFT.}
\label{fig:rho_t}
\end{figure}

We calculate the dynamics via KMC and DDFT for $L=9$. Since the total density $\rho$ grows during the deposition process, 
$Q(\rho)$ is an indirect way to visualize the time dependence of the nematic order $Q(t)$, but in contrast to $Q(t)$, $Q(\rho)$ can be directly compared with the equilibrium curve. In Fig.~\ref{fig:ddft_kmc} we compare KMC and DDFT (CORR) with varying degrees of growth; we employ both perpendicular  as well as isotropic deposition. 
There is very gratifying agreement between theory and simulation, although {with a small} deviation {only} in the isotropic--deposition case. 
This is highlighted when we plot the order parameter against the surface packing fraction $\eta$ in Fig.~\ref{fig:ddft_kmc_eta}.
The deviation of DDFT from KMC with isotropic deposition appears to amplify with long rods; compare these results to $L=5$ in Fig.~\ref{fig2}(b) 
with $\alpha=3$. This is likely a combination of effects: the density functional is {less precise} for longer rods~\cite{Oet16}, 
and the {packing fraction $\eta(=L\rho_{1 2}+\rho_3)$} is particularly sensitive to resulting errors in the number density 
$\rho_{1 2} = \rho_1 + \rho_2$; so, the error in $\eta$ scales with $L$. Apart from this, isotropic deposition with its random 
{insertion} of rod species $i=1,2$ appears to emphasize errors in the calculation of $\rho_{1 2}$. Fig.~\ref{fig:rho_t}(a) shows satisfactory correspondence between DDFT and KMC for total number density $\rho$ for both deposition types, while Fig.~\ref{fig:rho_t}(b) highlights the errors when observing $\rho_{1 2}$ alone.

The {explicit} evolution of observables in time  such as in Fig.~\ref{fig:rho_t} can be compared {directly} 
between DDFT and KMC if the kinetic rates, rather than their ratios, are matched explicitly. We set $\ktum = \frac{1}{2}\alpha_{i \rightarrow j} = \frac{1}{2}$, as well as $k_{\rm ins} = \ains$ and observe the evolution of number densities during growth.  
If we re--scale the  time variable with the flux rate $t^\star = \ains t$, the evolution of number densities can be compared 
for different growth rates. {Naturally, for decreasing flux rate these curves converge to a single curve, 
the quasi--equilibrium growth curve obtained by the solution of Eq.~(\ref{eq:drhoeq}). From Fig.~\ref{fig:rho_t}(a)
one sees that the time evolution of the total density is very well--described by the quasi--equilibrium curve for all
deposition rates. Since the quasi--equilibrium curve is essentially determined by the equation of state (through
$\mu(\rho_i)$), a measurement of $\rho(t)$ can be regarded as an effective measurement of the equation of state.
This is different for $\rho_{1 2}(t)$ (Fig.~\ref{fig:rho_t}(b)) where the results for the fastest deposition rate
deviate considerably both in shape and magnitude from the quasi--equilibrium curve.  } 

{One may compare these results to a very simple generalization of the Langmuir growth model. The latter is formulated for the
adsorption of isotropic particles, corresponding to our lattice model with perpendicular rods only. The insertion probability is proportional to
the free substrate area, i.e. the time development of the density is governed by $\dot\rho = \ains (1-\rho)$ with the
solution $\rho(t)= 1-\exp(-\ains t)= 1 - \exp(- t^\star)$. It describes our solution for perpendicular deposition reasonably well.
In the case of isotropic deposition, the assumptions of the insertion probability being proportional to the free substrate area and of having
no tumble processes lead to $\dot\rho = \ains (1-\eta)$ and $\dot Q=0$. The solution $\rho(t) =\gamma (1- \exp(-t^\star/\gamma))$
($\gamma=3/(2L+1)$)  differs grossly from our solution.}

\subsubsection{The case $\ku < \kd$: attractive substrates}

\begin{figure}
\centering
 \epsfig{file=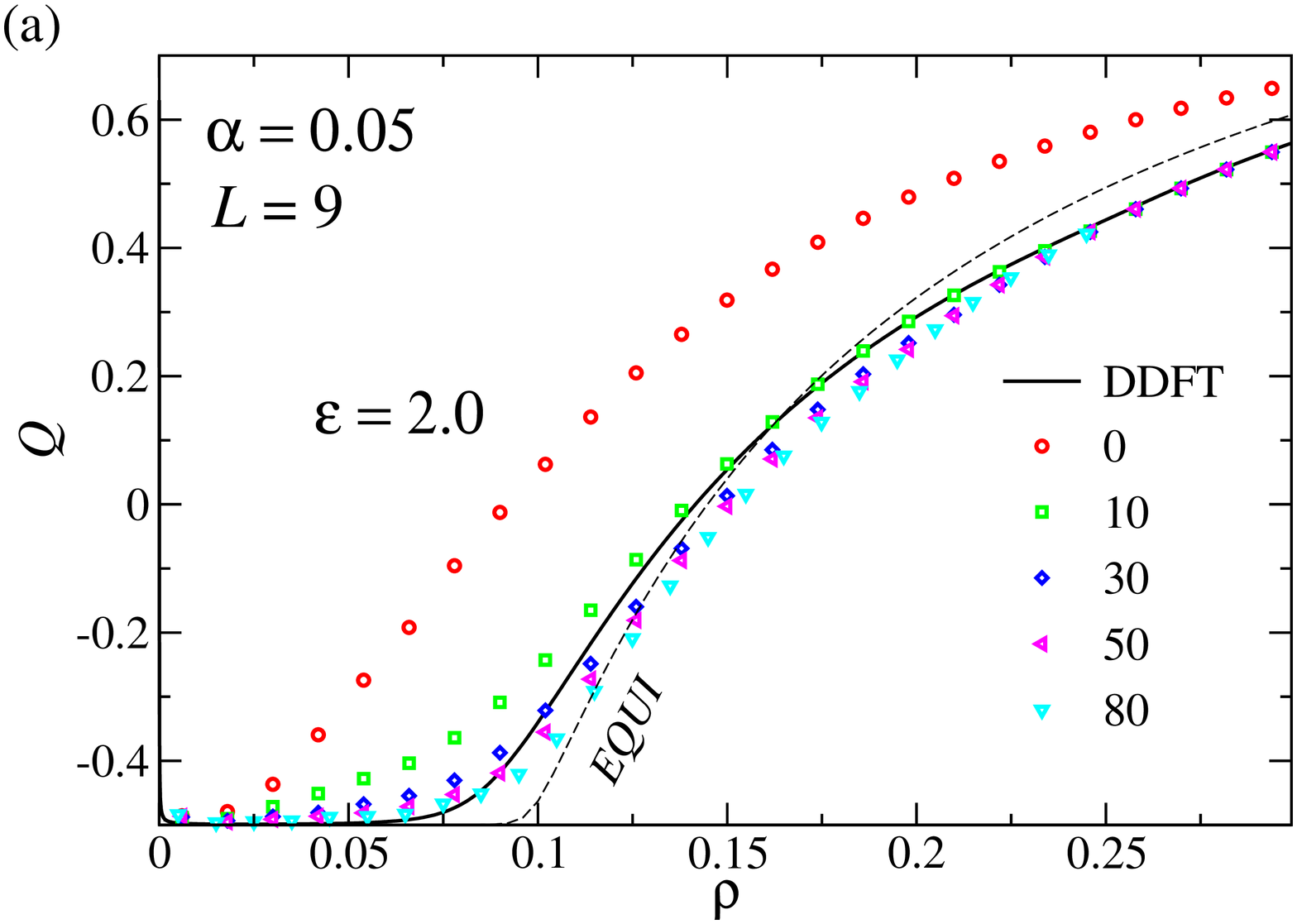,width=7.5cm}\hspace{3mm}
 \epsfig{file=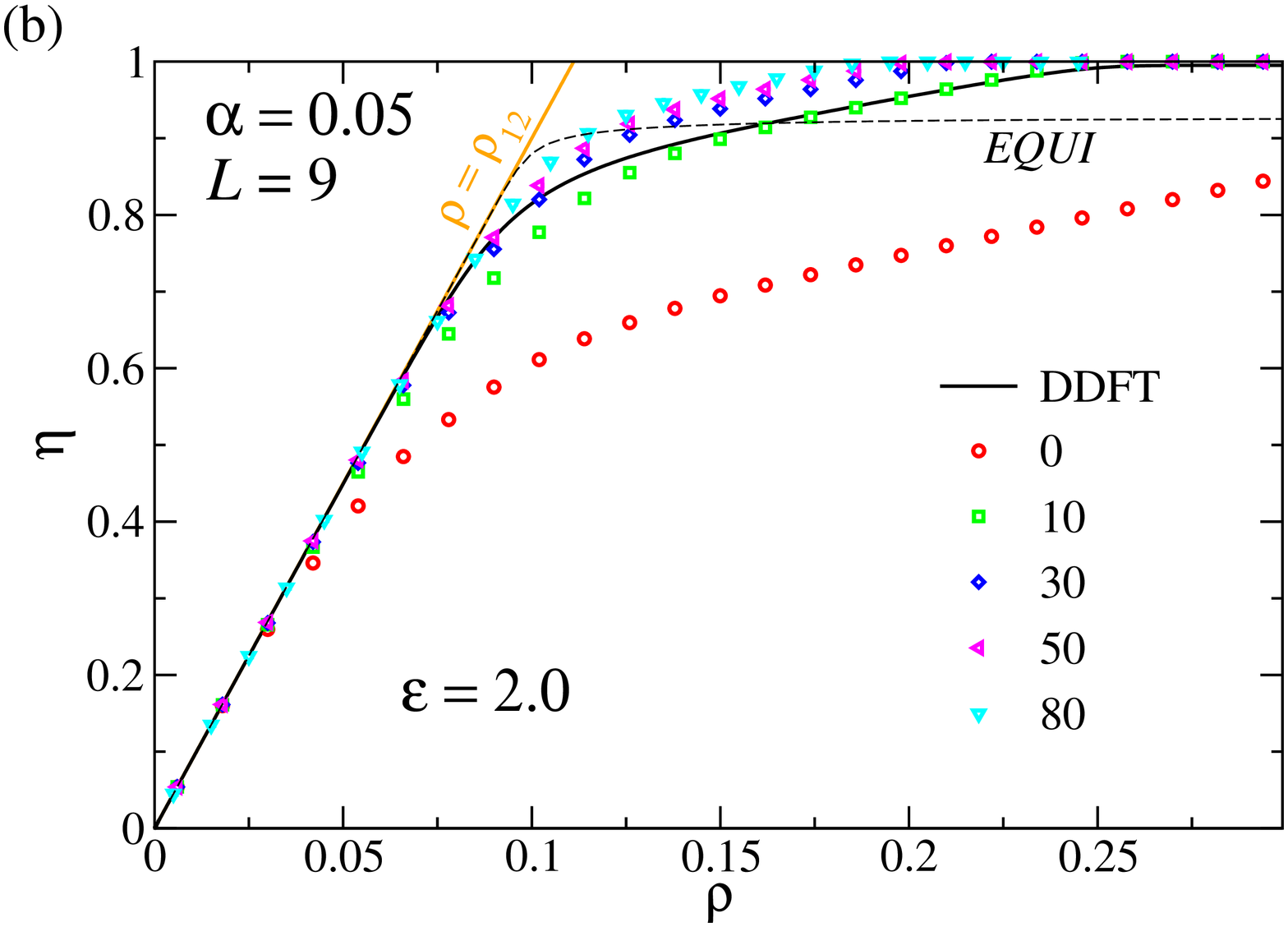,width=7.5cm}\\
 \vspace{1cm}
\epsfig{file=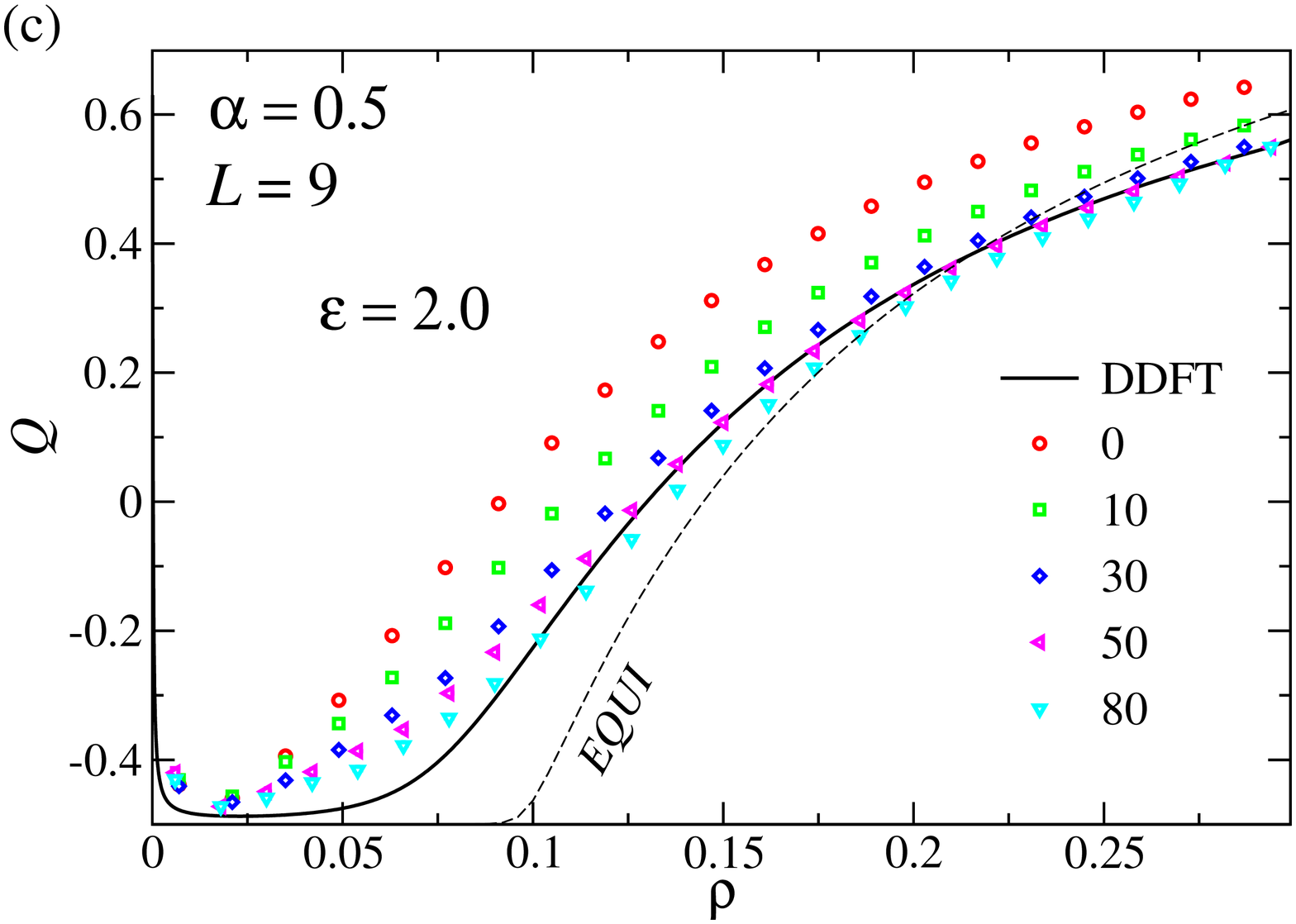,width=7.5cm}\hspace{3mm}
\raisebox{5mm}{\epsfig{file=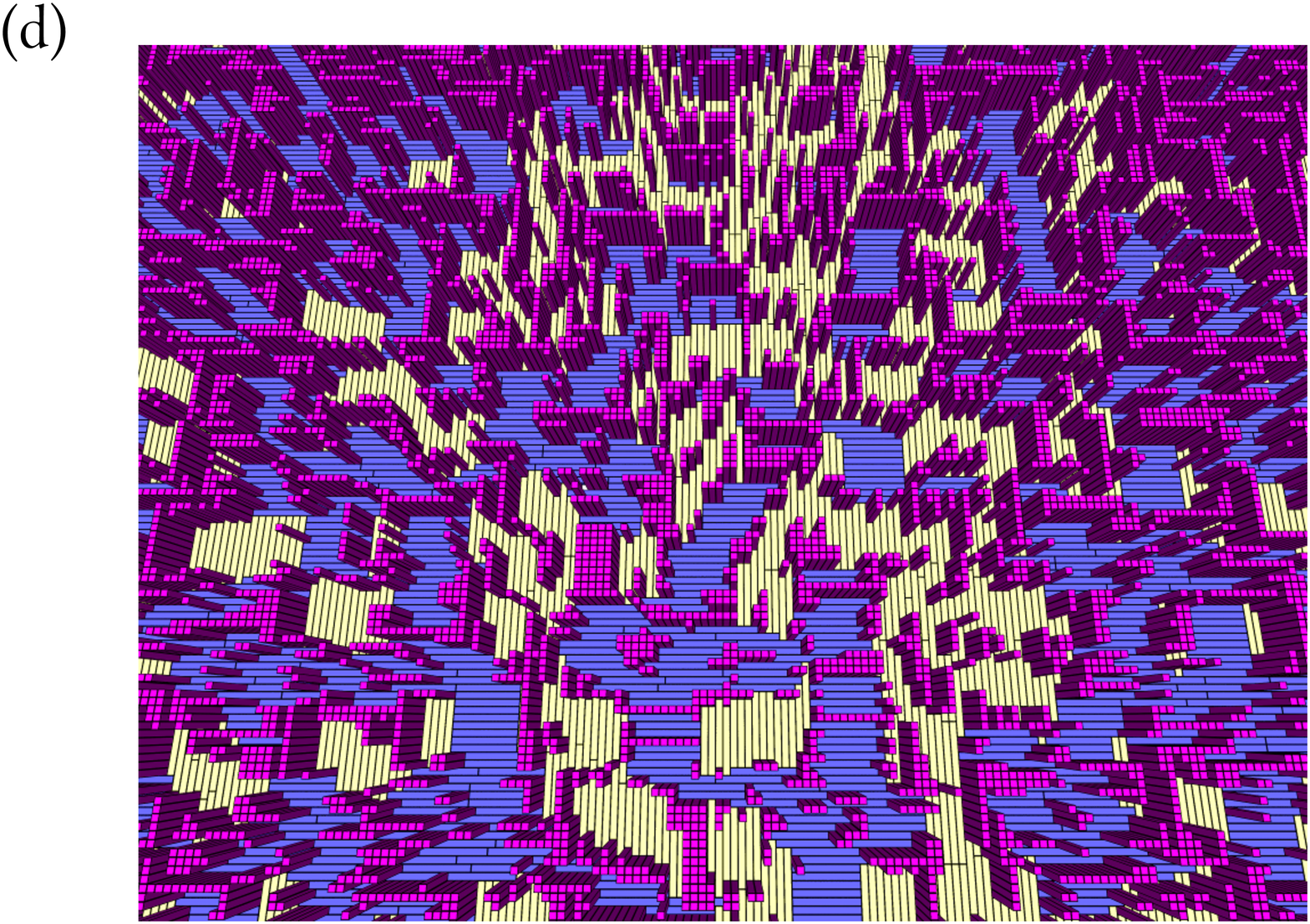,width=7.5cm}}
\caption{(color online) Growth of a monolayer of rods with lengths $L=9$ on attractive substrates under perpendicular deposition: 
Dependency of growth dynamics on translational diffusion. Indicated in the legend are values of the kinetic parameter 
{$\khopzero/(2 k_{\rm tum})$} (symbols with error bars). Shown additionally are DDFT calculations (black curves) as well as 
results from equilibrium--DFT (black dashed curves).  The substrate strength $\epsilon=2.0$. 
Growth with {comparably small rate $\alpha=0.05$} is represented (a) in the $(\rho,Q)$, and (b) in the $(\rho,\eta)$ plane, 
where {it is seen that} full packing is reached at relatively low densities. The limiting case of all rods lying on the substrate 
($\rho \equiv \rho_{1 2}$) is drawn in orange. (c) Same as (a), but for faster growth $\alpha=0.5$. Data points represent binned averages within single runs. (d) Illustration of a fully--packed configuration ($\eta=1$) at intermediate density. The color code is as in Fig.~\ref{fig:latticesetup}.}
\label{fig:attr}
\end{figure}

From the perspective of kinetics, the potential induces an additional energy barrier for the rods to stand up, where the 
activated dynamics is described by an Arrhenius law.\footnote{We model the effect of an energy barrier only on rotational 
motion---the substrate does not influence translational motion directly.}  The {corresponding rates employed follow the 
DDFT modeling (see Sec.~\ref{sec:ddft_vext}):
\bea
\ku = k_{\rm tum} e^{-(L-1)\epsilon/2}\\
\kd = k_{\rm tum} e^{(L-1)\epsilon/2}
\eea
}
Figure~\ref{fig:attr} shows the resulting dynamics from both KMC and DDFT calculations
{for the nematic order parameter $Q(\rho)$}.

A key feature distinguishes {the dynamics on attractive substrates from the one on neutral substrates}: 
the kinetic parameter $\khopzero$ contributing to translational 
diffusion  comes into play (compare Fig.~\ref{fig:khop}). {It appears that for large $\khopzero/k_{\rm tum}$  
the $Q(\rho)$--curves converge to a single one which is approximately described by the DDFT result.}
Now that tumbling moves are very rare events soon after a rod is introduced, this parameter alone controls local equilibration 
of the translational degrees of freedom.  This likely means strong configuration jamming occurs when {rod translations
cannot contribute to relaxation}. 
Figure~\ref{fig:attr}(b) shows that the surface becomes fully packed at $\eta=1$ at rather low densities $\rho$,  illustrated in Fig.~\ref{fig:attr}(d).  In an unusual change in character, the dynamics at full packing fraction {is dominated by} 
the rare events of rods standing up with (perpendicular) deposition taking place at the vacancies generated.

\newpage
\section{Continuous degrees of freedom: hard spherocylinders}
\label{sec:sphero}

{Similarly to our investigation of equilibrium monolayers in Ref.~\cite{Oet16}, we will explore
the possibility to match our lattice results in the dynamic case to corresponding results for a continuum
model with hard spherocylinders. One has to bear in mind, though, that the lattice model does not 
result from a systematic coarse--graining procedure applied to the continuum model. Rather, we attempt
to match basic dynamic parameters (i.e. characteristic microscopic times) and compare results.}

We have performed MC simulations off--lattice (with small displacement  and rotation moves) of hard spherocylinders 
{with length $\Lsph$, diameter $D_{\rm sph}$ and} 
aspect ratio $\kappa := \Lsph/D_{\rm sph}$ in the continuous 2D plane in a way analogous to those of~\cite{Oet16}. 
The minuscule MC moves induce pseudo--dynamics that on larger timescales {(where time is measured by the number of MC sweeps)} 
can be described by effective {translational and rotational} diffusion. 
\tr{As shown in Refs.~\cite{Berthier07,Patti12,Sanz10}, it  is possible to define a unique MC time scale being independent of the size of 
the MC change of any degree of freedom, and to relate such an MC time scale
to that of Brownian dynamics. As a matter of fact, in our case we only need to relate the MC time scale to that of KMC
for the lattice model.}
Apart from the Brownian translational and rotational motion, the continuum model also includes an external flux for 
introducing (depositing) rods into the system. 
To compare growth between the lattice and continuum models it is necessary to map the {characteristic times} of their 
microscopic kinetics. {In the continuum model, these are $\{1/(F D_{\rm sph}^2), \tau_{\rm diff}, \tau_{\rm rot} \}$
(with $F$ denoting the influx or deposition rate per area and a unit area is chosen by $D_{\rm sph}^2$, $\tau_{\rm diff}$ the translational self--diffusion time and
$\tau_{\rm rot}$ a rotational relaxation time)---these must be matched to the three times
$\{ 1/(\kins{} u^2), 1/k_{\rm tum}, 1/k^0_{\rm hop} \}$ in the lattice model ($u=1$).}
We discuss this matching procedure generally, at first, before applying it to two very different growth models in sections 
\ref{sec:appearing_rods} and \ref{sec:appearing_rods}. As in our previous work, the order parameter {used in the continuum
model} is the largest eigenvalue, $Q_{\rm nem}$, of the nematic order tensor.

\subsection{Basics of matching to lattice model}

{To avoid additional complications due to correlations}, we will perform the matching for the three timescales {in the case of a dilute monolayer, i.e.
for the initial stage of film growth}.
Furthermore, we address only the case of neutral substrates.

\subsubsection{Translational diffusion}
\label{sec:trans_diffusion}

{In the continuum model, the translational self--diffusion time over a distance $D_{\rm sph}$ is given by
\bea
  \tau_{\rm diff} = \frac{D_{\rm sph}^2}{\DtwoDcont}\;.
\eea
The equivalent time in the lattice model would be the translational self--diffusion time over a distance $u$ (lattice unit).
Matching these gives
\bea
  \frac{D_{\rm sph}^2}{\DtwoDcont} = \frac{\unit^2}{\DtwoDlatt} \;.
\eea
For a dilute system of rods in the lattice model, the translational diffusion constant is given by
${\DtwoDlatt}/{{\unit}^2} =\khopzero + \khop(\kd,\ku)$ (see Eqs.~(\refeq{eq:D_latt}),(\ref{eq:khop}) and App.~\ref{app:derivation_D2Dlatt}),
where $\khopzero$ accounts for the explicit translational move and $\khop$ is due to the tumbling move.
For vanishing substrate potentials ($\kd=\ku\equiv\ktum$) the contribution from tumbling becomes }
\bea
\label{eq:khopsymmktum}
\left.\khop\right\vert_{\rm \kd=\ku} =    \frac{2}{3}  \frac{(L-1)^2}{4}\ktum\;.  
\eea
This value is fixed for a given tumbling rate and rod--length $L$.

In the continuum model (with given translational and rotational moves), 
we measure $\DtwoDcont$ directly in a separate simulation where rods behave like an ideal gas
{and the diffusion constant is extracted from the slope of the mean--square displacement of a rod versus  simulated time. 
In this way, the diffusion rate from the translational move in the lattice model can be fixed to
\bea
  \khopzero = \DtwoDcont/\Dsph^2 - \khop(\ktum)\;.
 \label{eq:khopmatch}
\eea
The tumbling rate $\ktum$ entering the equation above is fixed by a concrete rotational relaxation time (see below).} 

{However, in Sec.~\ref{sec:matchddft} we showed that the dynamics does not depend on $\khopzero$ in the lattice model as long as there is no external potential. 
As our investigation is restricted to this condition, the particular value of $\khopzero$ does not play any role for evolution of the total density and
the orientational order and is put to zero. 
For a closer investigation of the case of finite substrate potential, one would need to take  
the condition in Eq.~(\ref{eq:khopmatch}) into account.}

\subsubsection{Rotational relaxation}
\label{sec:rotations}

{
In the continuum model, the rotational relaxation time $\tau_{\rm rot}$ can be defined by the relaxation time for nematic order, i.e.
the characteristic decay time in the autocorrelation function $\langle Q_{\rm nem}(t) Q_{\rm nem}(0) \rangle$ in a dilute system.
We obtain this decay time by recording the autocorrelation function in a system of spherocylinders with no interactions
and fitting it to an exponential ($\propto\exp(-t/\tau_{\rm rot})$). 
 }

{
In the lattice model, the corresponding autocorrelation function $\langle Q(t) Q(0) \rangle$ can be obtained analytically
in the ideal--gas limit and the characteristic decay time is $\tau_{\rm rot}^{\rm latt} = (6 \ktum)^{-1}$.
For given rotational moves in the continuum simulation, the tumbling rate $\ktum$ is determined by matching these times.
} 

{
Instead of using the characteristic decay time of orientational (nematic) order, one might think of matching the rates for a transition
from a standing--up to a lying--down rod. In the lattice model, this would be affected by 4 possible discrete moves, each with
rate $\ktum$. Thus, this transition time is $1 /({4\ktum})$. In the continuum model, this transition time
would  be the first passage time for a rotation from standing to lying, which we also determined in a simulation with
ideal spherocylinders. However, this first passage time is about 100 times larger than the decay time for nematic order.
Through the comparison of lattice and continuum results (see below) we find that matching the decay time for nematic order
is sensible and matching the first passage time leads to grossly different results.
The reason is that in the autocorrelation function measures a continuous change of order. A certain change $\Delta Q$ in the lattice model
comes about by a fraction of rods reorienting in the lattice model, whereas for a corresponding change $\Delta Q_{\rm nem}$
 in the continuum model, the spherocylinders need (on average) to reorient the same amount. The corresponding 
time needed is much smaller than the first passage time for a rotation from standing to lying.
}

{\subsubsection{Deposition time and growth parameter}}
   
{
The characteristic time for deposition on the unit area for a dilute system does not depend on the diffusional properties and is simply
given by $1/\kins{}$ (lattice, $u=1$) and $1/(F D_{\rm sph}^2)$ (continuum). Hence the growth parameter $\alpha$ must be matched between lattice and continuum
in the following way:
\bea
  \alpha = \frac{\kins{}}{2 \ktum} = \kins{} (3 \tau_{\rm rot}^{\rm latt}) = (F D_{\rm sph}^2) (3 \tau_{\rm rot}^{\rm cont}) \;.
\eea 
}


\subsection{Model I: Deposition as random `appearance' of rods}
\label{sec:appearing_rods}

  In this model, the midpoints of the hard rods are constrained  to a continuous 2D plane of size $\lbox^2 = 200\times200\Dsph^2$ 
with periodic boundary conditions.  They rotate freely and  diffuse along the substrate via small MC moves as to approximate Brownian dynamics.
{Rotational moves are performed as described in Ref.~\cite{FrenkelSmit}.}
New rods are introduced to the monolayer (they `appear') with a global rate {$\rins = F\,\lbox^2$}.
As in the lattice model, hard--core repulsion between the rods means an attempt at inserting a rod at some position and with a certain orientation is 
\emph{rejected} if it overlaps with another. {Time progresses also for these unsuccessful deposition attempts}, causing the number density of rods 
to depend on time in a monotonic, but non--linear way, see Fig.~\ref{fig:rho_t_cont} below. 
As in the lattice model, we employ two deposition conditions: one where rods are deposited in a vertical orientation and another in random, isotropically--distributed orientations. 
Results are presented in Sec.~\ref{sec:resultsI}, whereby the parameters are indicated below. 
\begin{figure}
  \includegraphics[width=0.5\linewidth]{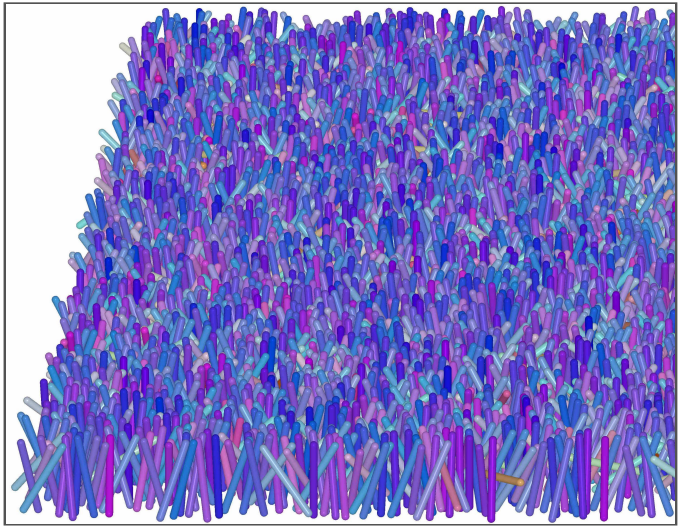}
  \caption{(color online) Illustration of continuum Model I for a monolayer of hard spherocylinders.}
\end{figure}

 
Our investigations are performed for rods of length $L=9$ (lattice) and aspect ratio $\kappa=8$ (continuum) since the spherocylinders have total 
length $\Lsph+\Dsph$. We note that {results for $\kappa=9$ are very similar and will not be shown}.


\subsection{Model II: Deposition as sedimentation {caused by a constant force (``gravity'')}}
\label{sec:depositing_rods}

 In this model, the hard rods move in 3D space via small rotations about their midpoints and translation moves in 3D. 
They fall onto a square--well attractive substrate (well ${\rm depth} = 50\kt$, ${\rm width }=0.05 \Dsph $) in a box with 
periodic boundary conditions {in the substrate plane}. The attractive substrate is not of the sort described in 
Sec.~\ref{sec:ddft_vext}---rather, it acts as an ``adherent" {where the rod experiences the well (with the orientation--independent depth) 
only if
the surface--to--surface distance to the substrate is less than the width of the square--well potential. 
Thus it serves}  as a strong barrier against rods desorbing. Rods diffuse and rotate by the same MC moves as in Model I, even though, 
now, midpoints are unconstrained above the substrate and diffusion moves are generated in 3D. Rods are generated with  
random positions and orientations at the top of the box ($\lbox^2\times l_z = 50\times 50\times 100\Dsph^3$) and inserted at a fixed rate 
{$\rins$}; hence we only investigate isotropic deposition. They `fall' to the bottom of the substrate under an artificial 
gravitational force $g$. 
In order to disentangle gravity and {the adhesive} substrate potential, we switch--off the gravity when the $\hat z$--coordinate of the 
rod midpoint is less than half a rod--length  
$(\Lsph+\Dsph)/2$,  where \z{} is normal to the substrate. 
This model qualifies for 3D multilayer growth, emulating thin film growth with, say,  OMBD more closely; however,  
we investigate only  exemplary cases as the 3D nature of this model deviates significantly from the  lattice system in focus. 

 
{In order to match the characteristic deposition time, we need to determine the deposition rate per unit area (flux) $F$.
Our MC pseudo--dynamics result in a net drift of the rods towards the substrate with velocity $v = \Gamma g$, where the
mobility $\Gamma$ is determined by the translational diffusion constant in 3D $\DthreeDcont$ through $\Gamma = \DthreeDcont/\kt$.
The flux is then given by $F = \rhobulk v = \DthreeDcont/\kt \rhobulk g$, where $\rhobulk$ is the 3D number density of rods well--above the substrate.
In the simulations, we fix $\rhobulk = 10^{-4}/\Dsph^3$ and measure $\DthreeDcont$ through the slope in the mean--squared displacement
vs. time. Matching the flux between lattice and continuum is achieved by appropriately choosing $g$. 
} 
 

{For matching the self--diffusion time, we}
measure the diffusion rate $\DtwoDcont$ (see Sec.~\ref{sec:trans_diffusion}), but, this time for an ideal gas of rods adhering to the substrate. 
Note that although the MC moves for translations continue in 3D, the substrate potential almost always causes a Metropolis rejection for a 
move escaping the potential barrier. As this barrier is very thin ($0.05\Dsph$), the restricted 3D diffusion is effectively 2D diffusion. 
Similarly, we match the rotational relaxation time by measuring the autocorrelation function for nematic order as described
in Sec.~\ref{sec:rotations}. {We note that orientational diffusion of rods in model II arises from a combination of
midpoint rotation moves and vertical moves since the rods must remain close to the adhering substrate. This leads to
an autocorrelation of $Q_{\rm nem}$ nondescribable by a single exponential. For determining $\tau_{\rm rot}$, we fitted
the initial decay.} 


Our monolayer orientational observables are calculated strictly for rods adhering to the substrate, {with} the number density in the 
monolayer denoted by $\rho_{\rm subs}$. We additionally analyze the total density across the \z--direction, 
in particular as a `second layer' may form.

\section{Growth Results}
\label{sec:results}

\subsection{Model I}
\label{sec:resultsI}
\begin{figure}[htb]
\epsfig{file=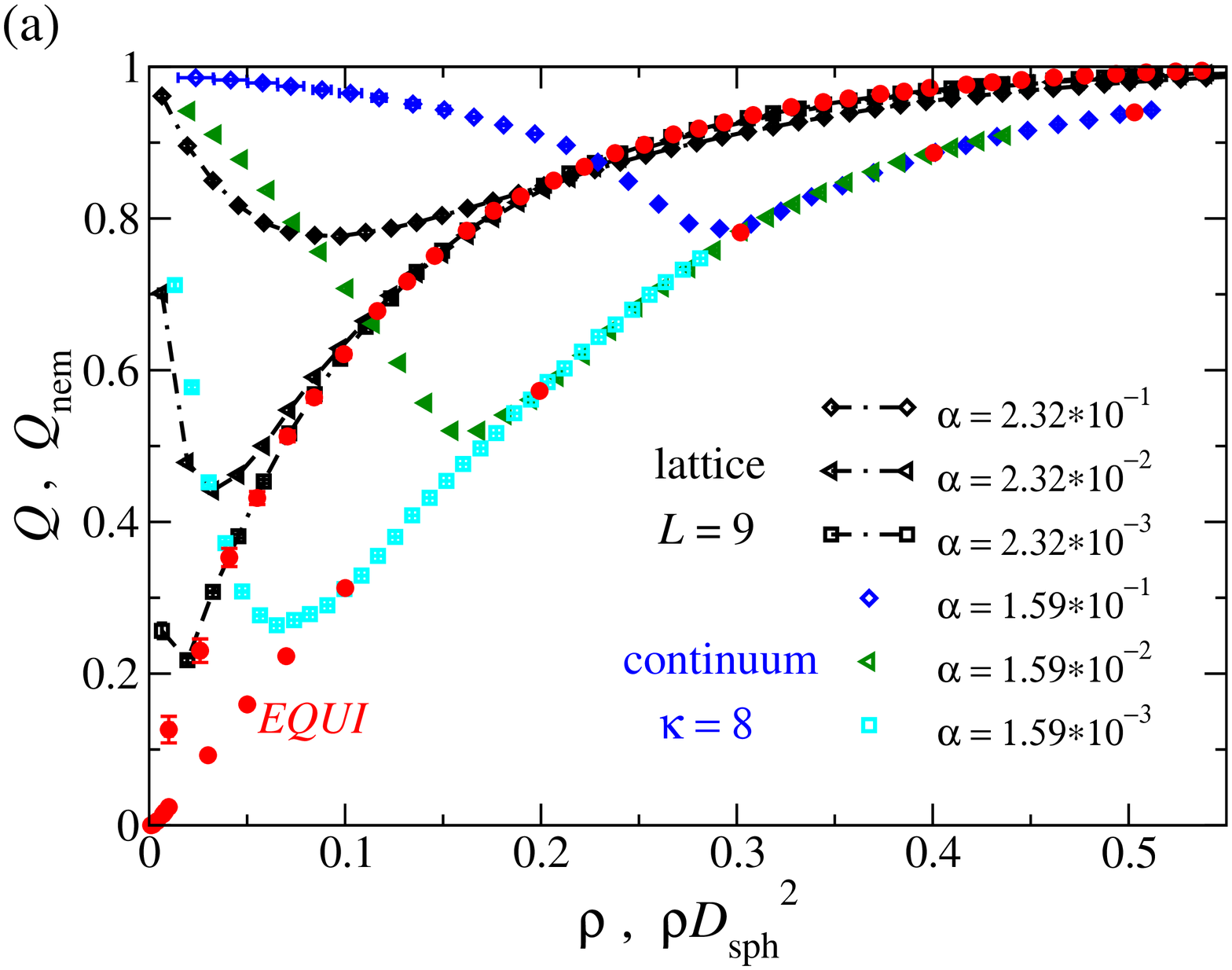,width=7.5cm}\hspace{3mm}
\epsfig{file=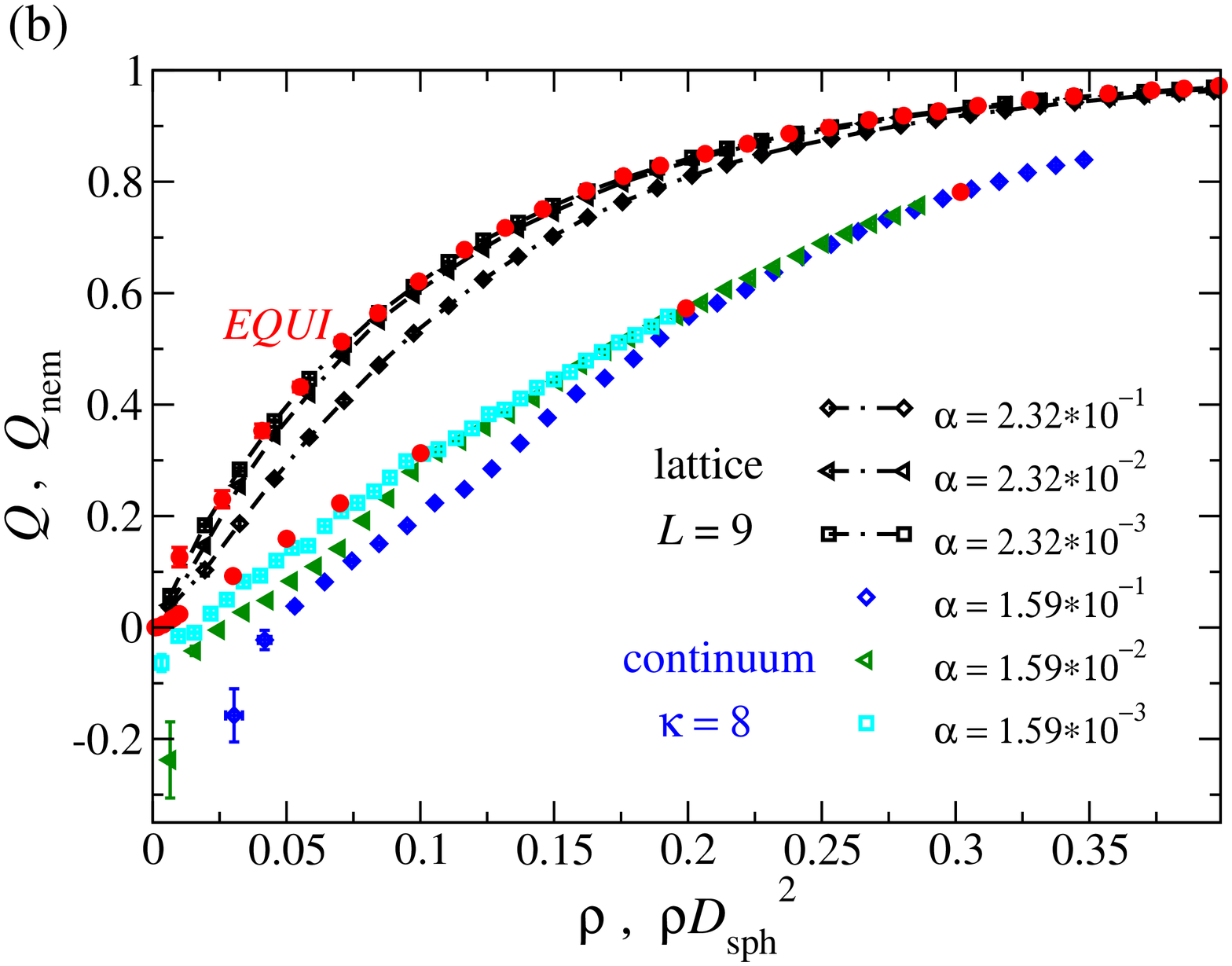,width=7.5cm}
\caption{(color online) Growth of a monolayer of hard spherocylinders ($\kappa=8$) using Model I (appearing rods) represented in the $(\rho,Q_{\rm nem})$ 
plane: comparison to lattice model (black)  with matched kinetics. Monolayers are grown with perpendicular deposition (a) and 
isotropic deposition (b). 
Red data points correspond to thermodynamic equilibrium in the lattice (steep curve) and continuum (shallow curve) models. 
}
\label{fig:Q_rho_modelI}
\end{figure}

In Model I, where new rods `appear' within the monolayer, one might expect the {evolution} of the order parameters  $Q$ (lattice) 
and $Q_{\rm nem}$ (continuum) {with the total} number density $\rho$ to look similar to Fig.~\ref{fig:ddft_kmc}. 
Indeed this is what we find in Fig.~\ref{fig:Q_rho_modelI}, where we varied the growth parameters $\alpha$ {over two decades}. 
All continuum data are running averages over a single run, and equilibrium data points for the lattice model are obtained via Grand Canonical 
Monte Carlo simulations \cite{Oet16}. Most striking in the figure is the similar form of the curves for growth under perpendicular deposition
{with respect to the equilibrium curve (Fig.~\ref{fig:Q_rho_modelI}(a))}. In both models, {the downward dip of the order parameter 
and subsequent approach to the equilibrium curve happens at about the same value of $Q$ ($Q_{\rm nem})$, but it is shifted to higher 
densities in the continuum.}
\tr{For dilute systems, the shift in densities can be attributed to the different two-body excluded volumes
in the lattice and the continuum model. If the lattice densities are multiplied by the ratio of the volumes, which
is approximately given by \cite{Oet16} $\frac{L^2 + L -2}{9\cdot 0.45 L}$ ($\approx 2.5$ for $L=9$), 
the agreement between the lattice and continuum models is quite good for continuum densities
$\rho D_{\rm sph}^2 < 0.2$, yet differences remain for higher densities.}

\begin{figure}[htb]
\epsfig{file=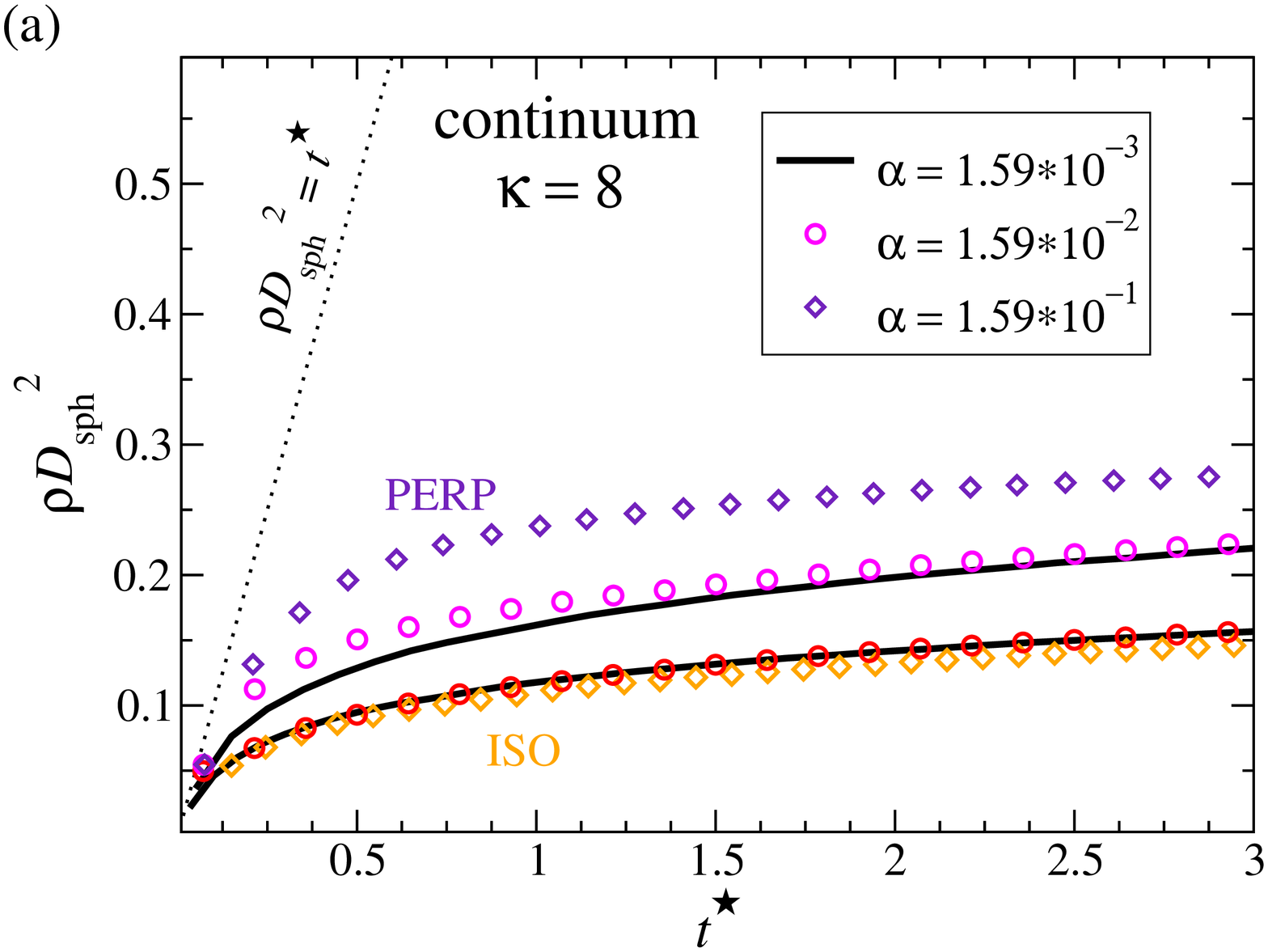,width=7.5cm}\hspace{3mm}
\epsfig{file=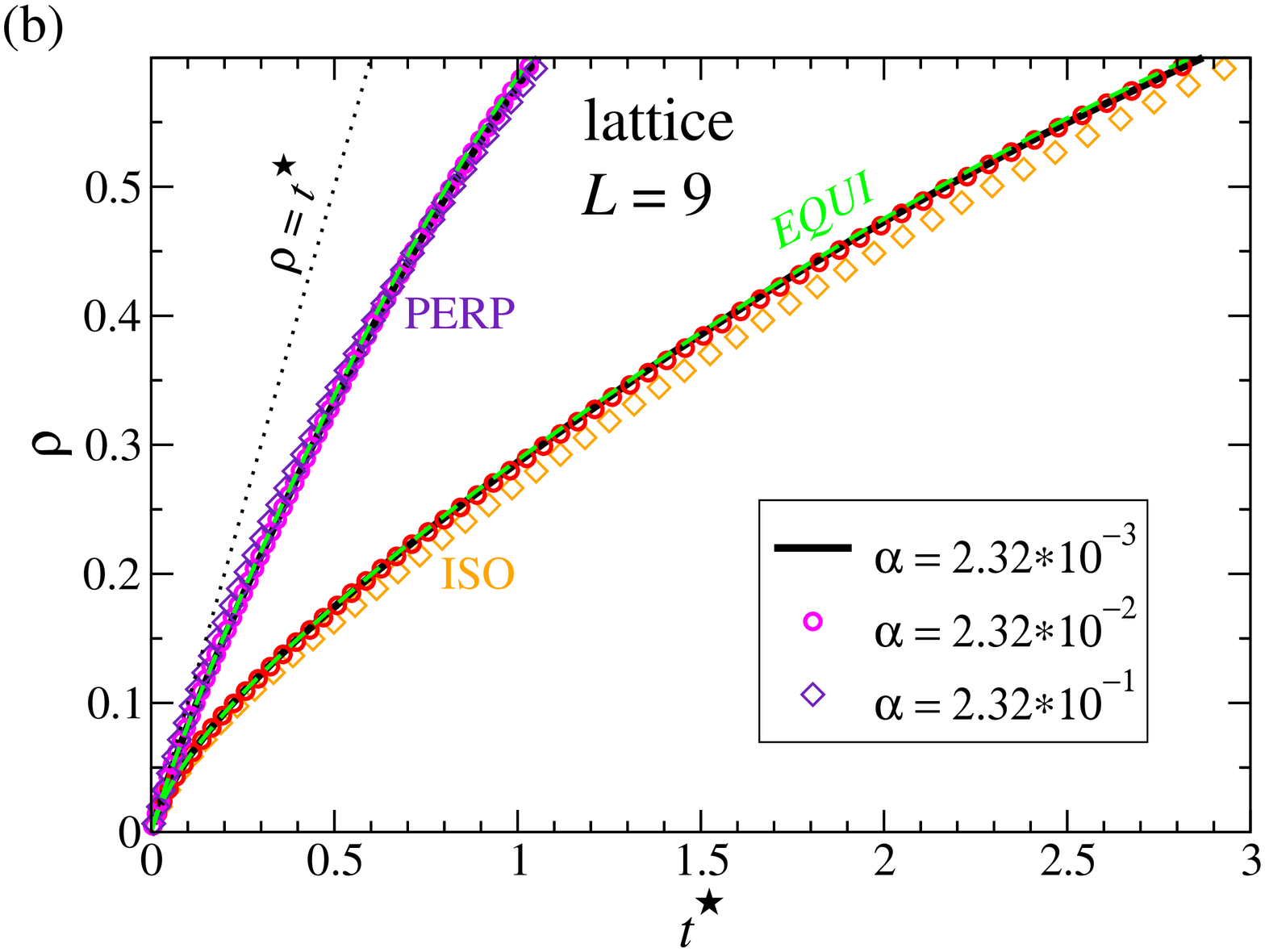,width=7.5cm}
\caption{(color online) Evolution with re--scaled time  $t^{\star}$ of number densities $\rho$ 
during monolayer growth in continuum Model I and lattice. Perpendicular deposition (PERP) is shown in purple, 
while  isotropic deposition (ISO) is shown in orange. {Same symbol shapes/line--style refer to the same growth parameter $\alpha$.} (a) Continuum, $\kappa=8$. (b) Lattice, $L=9$. 
The green dashed curves (EQUI) correspond to quasi--equilibrium growth calculated with DFT. 
{Deposition of ideal--gas rods (dotted lines) describes the initial slope in $\rho(t^\star)$.}
}
\label{fig:rho_t_cont}
\end{figure}

Fig.~\ref{fig:rho_t_cont} displays the dynamics of the total number density $\rho$ versus the {re--scaled time 
$t^\star=\kins{} t$ (lattice) and $t^\star=(F D_{\rm sph}^2)  t$ (continuum), respectively.
The quasi--equilibrium growth curve for the lattice model (see Sec.~\ref{sec:quasieq}) is also shown in 
Fig.~\ref{fig:rho_t_cont}(b). 
For the continuum model {with isotropic deposition} (Fig.~\ref{fig:rho_t_cont}(a)), there is little variation of $\rho(t^\star)$ with $\alpha$
(as in the lattice model), and the results seem to be well--described by a quasi--equilibrium growth curve,
which would be attained for  $\alpha\to 0$. {For perpendicular deposition, the results for the highest growth rate ($\alpha\approx 0.16$)
are different from those for the two lower growth rates, but they converge for later times $t^\star \gtrsim 20$.}
We point out a strong difference when comparing these growth curves for the lattice and continuum: In the continuum model, the density increases only very slowly beyond the dilute limit ($t^\star \gtrsim 0.1$).
Since the quasi--equilibrium growth curve is determined only by the equation of state (through $\mu(\rho(\theta))$ where
$\theta$ is the polar angle), this indicates the equations of state in the lattice and continuum model, respectively, 
are very different already for moderate densities. The continuum equation of state for the full density range
is not known. In Ref.~\cite{Oet16} we only analyzed a virial expansion up to second order. Already at this order we found a different
scaling of the second virial coefficient: it is $\propto L_{\rm sph} D_{\rm sph}$ for the
continuum model and $\propto L^2$ for the lattice model.  } 
\newpage

\subsection{Model II}
\label{sec:resultsII}

\begin{figure}[htb]
\epsfig{file=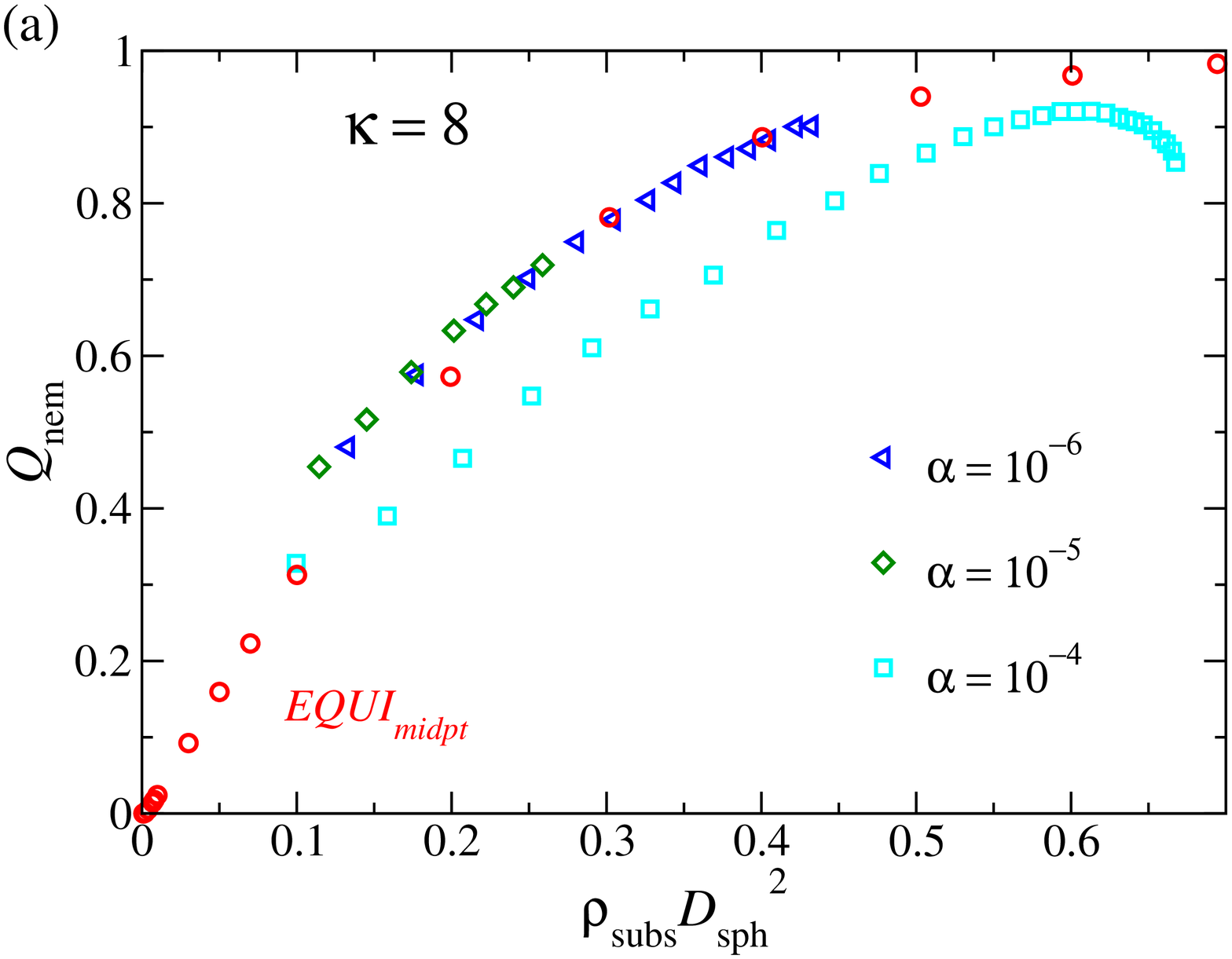,width=7.5cm}\hspace{3mm}
\epsfig{file=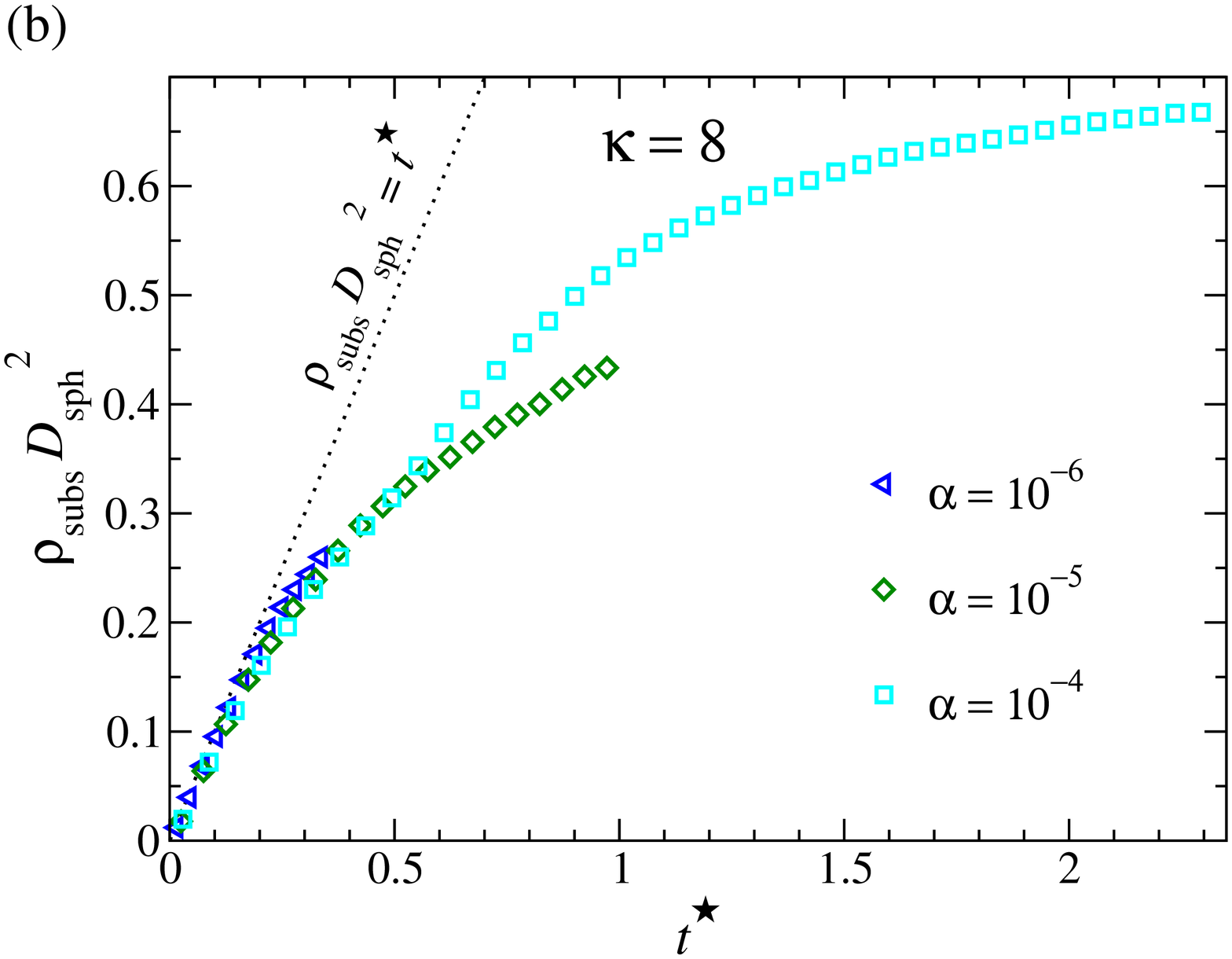,width=7.5cm}\\
\vspace{6.5mm}
\raisebox{2mm}{\epsfig{file=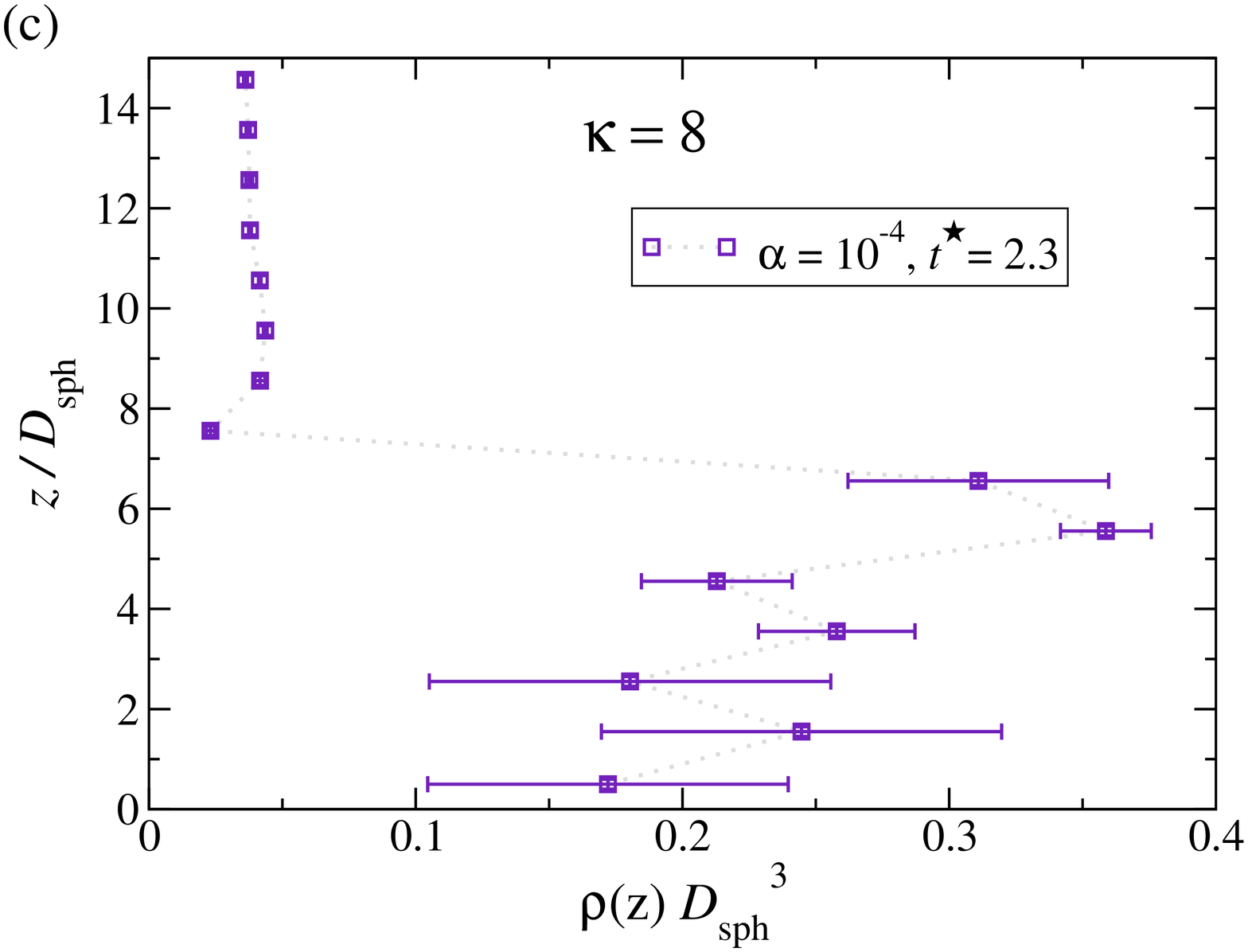,width=7.5cm}}
\caption{(color online) {(a) Nematic order vs. density in the first grown layer of hard spherocylinders (aspect ratio $\kappa=8$) using Model II 
(rod sedimentation)  for different values of $\alpha$. 
 Red data points correspond to thermodynamic equilibrium in the continuum model with fixed midpoints. 
 (b) Evolution of the density in the first grown layer $\rho_{\rm subs}$ with re--scaled time $t^\star$. 
 The deposition curve for ideal gas particles (dotted lines) means that (on average) all drifting
 particles reaching the substrate will stick to it. 
 (c) Height above substrate versus 3D rod density  
 for a growth parameter $\alpha=10^{-4}$ and $t^\star=2.3$, corresponding to the last point in (a). 
 The increased density for $z/D_{\rm sph}>4.5$ signals the formation of a disordered second layer.} }
\label{fig:Q_rho_modelII}
\end{figure}

Figure~\ref{fig:Q_rho_modelII}(a) displays growth of the monolayer in the $(\rho_{\rm subs}\Dsph^2,Q_{\rm nem})$ plane, where 
$Q_{\rm nem}$ is calculated for all rods adhering to the substrate (i.e. those contributing to $\rho_{\rm subs}$). 
The equilibrium curve {shown} corresponds to that of rods with fixed midpoints, i.e. the system in Model I. 
{For the two smaller growth rates {($\alpha =10^{-6}$ and $10^{-5}$)} the nematic order in the monolayer is close to the
equilibrium curve, similar to Model I.} On the other hand, faster growth (cyan squares, {$\alpha=10^{-4}$}) 
{shows different behavior}: 
the nematic order is {noticeably} lower, an effect also seen in the isotropic--deposition growth of 
model I (Fig.~\ref{fig:Q_rho_modelI}(b)). {Furthermore, at higher densities the monolayer does not converge to a fully--ordered
state. $Q_{\rm nem}$ drops, instead. This is an effect of particles accumulating on top of the first layer.} 

{ Figure~\ref{fig:Q_rho_modelII}(b) shows $\rho(t^\star)$ for model II. The initial, linear behavior characteristic
of deposition on a dilute layer is similar to model I; however, for $\rho_\text{subs} \Dsph^2 \gtrsim 0.2$
significant deviations appear. There, growth in model I becomes very slow (see Fig.~\ref{fig:rho_t_cont}(a), ISO curves).
In model II, new rods increasingly `hover' above the monolayer, breaking the single--layer assumption and leading to
enhanced adsorption in the first layer. Convergence to a quasi--equilibrium growth curve for low $\alpha$ can be seen only up to
$\rho_\text{subs} \Dsph^2 \approx 0.3$.} 

{
In the monolayer growth regime, two major differences between model I and II can be observed.
(i) The curve $\rho(t^\star)$ in model I quickly bends over and stays near $\rho D_{\rm sph}^2=0.15$ for a long time.
This is not so in model II. Apparently almost all the
rods that are in the vicinity of the substrate reach it within a short time--period.
This happens since rods 
 diffuse around in the vicinity of the substrate and finally reach it after multiple ``attempts''. The fluxes employed are small so that diffusion
 is a reasonably fast process. (For the lowest $\alpha$, the first rods reach the substrate not by the sedimentation
 drift but by bulk 3D diffusion).
(ii) The growth parameters used to study model II are well in the quasi--equilibrium growth regime for model I. Nevertheless, we see these values of $\alpha$ generating clearly non--equilibrium behavior that also differ significantly in character to model I. We conjecture that an effective $\alpha$ for model II is actually higher than reported owing to the aforementioned bulk 3D diffusion.
}

\begin{figure}
 \epsfig{file=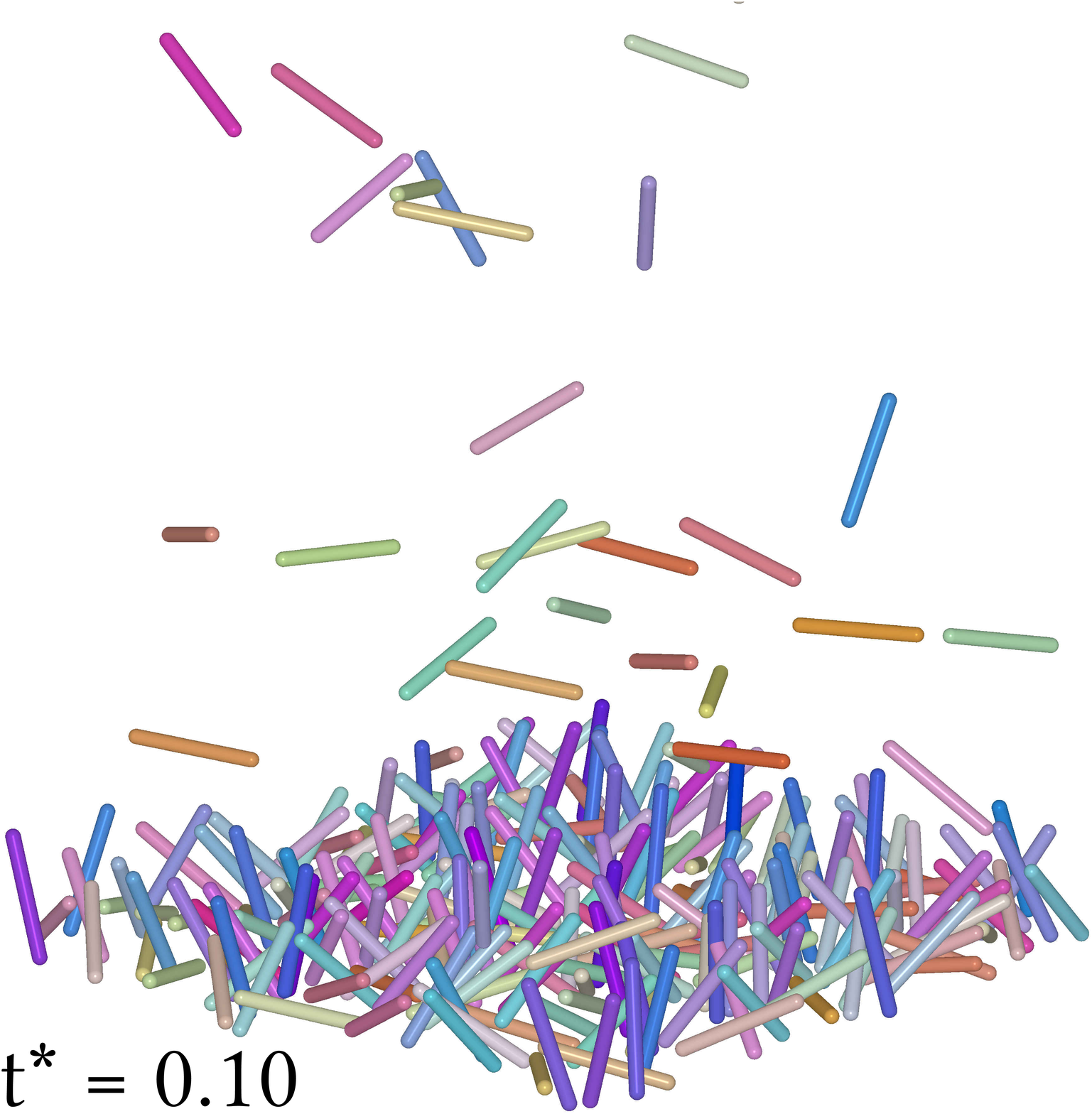,width=0.3\linewidth}\hspace{1mm}
 \epsfig{file=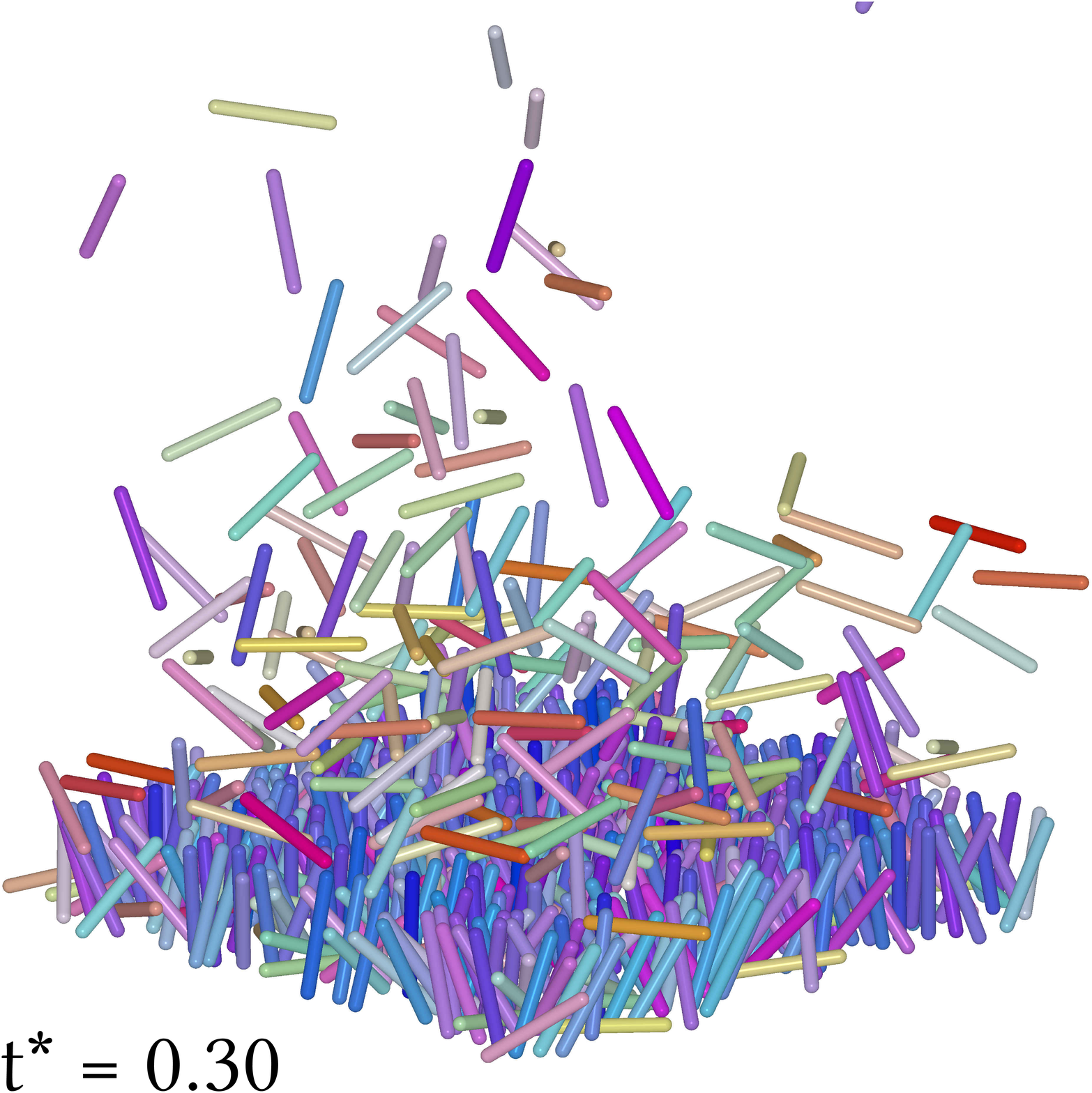,width=0.3\linewidth}\hspace{1mm}
  \epsfig{file=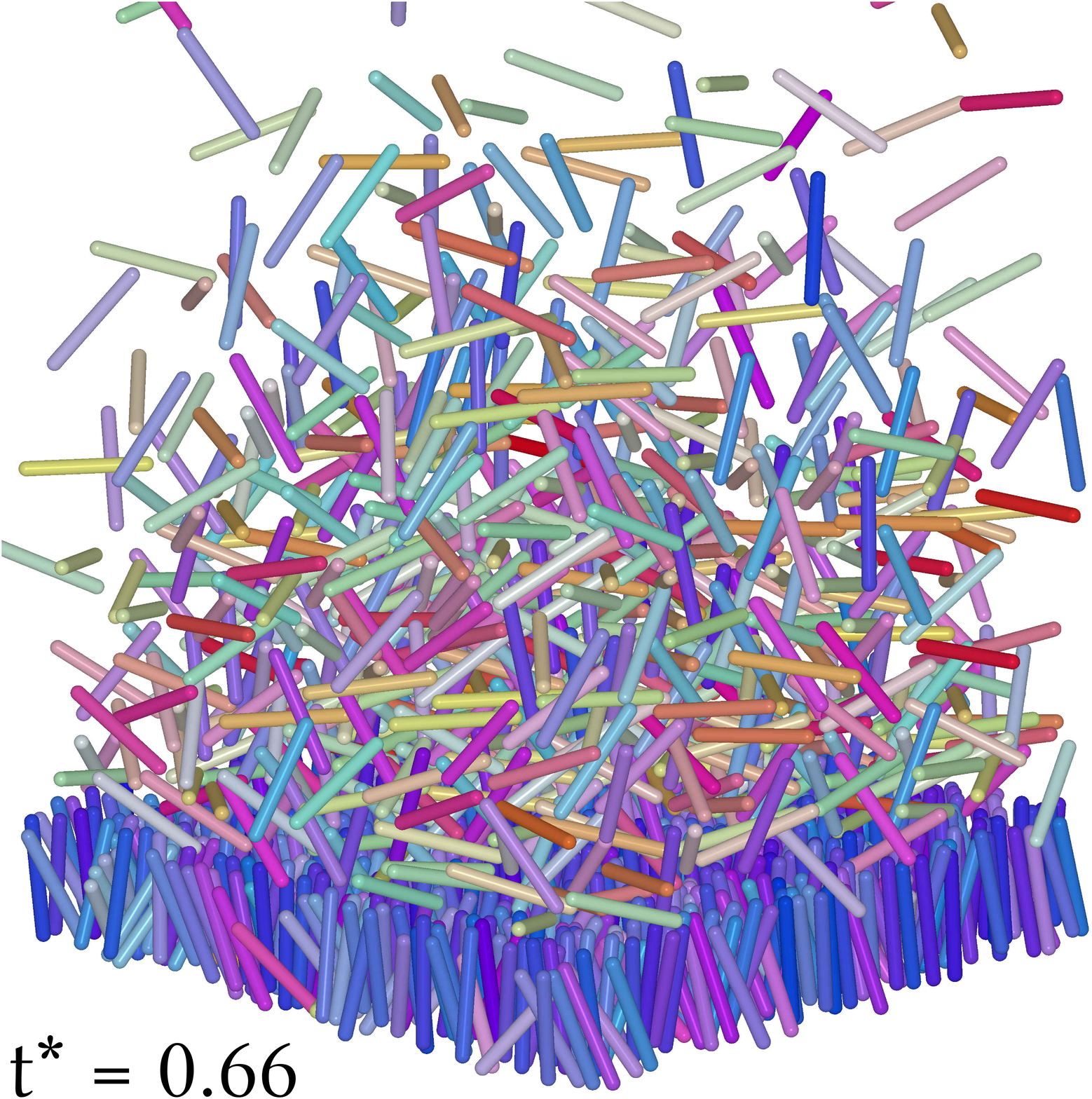,width=0.3\linewidth}
  \caption{(color online) Snapshots of growth in Model II for aspect ratio $\kappa=8$ and {growth parameter $\alpha=10^{-5}$}.}
  \label{fig:snapshots_modelII}
\end{figure}

In the regime past the monolayer, we comment on a few preliminary findings: 
{As aforementioned, in the vicinity of} reduced densities of 0.6 {in the monolayer}, the nematic order drops due to a population of rods building up above the monolayer, jamming up space for rods in the first layer. {An exemplary} distribution of rods versus vertical height
{for this regime is} shown in Figure~\ref{fig:Q_rho_modelII}(c). {Rods in the monolayer
contribute to the measured density $\rho(z)$ only up to $z/D_{\rm sph}=4.5$; thus, increased density for larger $z$
belongs to a second layer. This second layer is very disordered as corresponding snapshots suggest
(see Fig.~\ref{fig:snapshots_modelII}).}

\newpage
\section{Summary and Outlook}
\label{sec:summary}

{We have conducted a study of monolayer growth in hard rod models using dynamic lattice DFT,
lattice KMC simulations and continuum simulations with diffusive dynamics. 
The hard rod models employed do not aim to describe a specific system but rather emphasize the steric effects which can
occur when looking at, e.g., Langmuir monolayers or the initial, sub--monolayer stage of film growth with anisotropic molecules.
The nematic order $Q$ in the monolayer is
due to entropy alone, and its growth with density or time is clearly dominated by the equilibrium properties of
the monolayer. For a wide range of growth rates, the time evolution of total density $\rho$ in the monolayer is in fact described 
by a quasi--equilibrium curve in which the monolayer equation of state enters. Dynamic effects (deviations from
quasi--equilibrium) are more pronounced when monitoring $Q(\rho)$ or, moreso, $Q(\eta)$, where $\eta$ is the 
packing fraction in the monolayer.}

{For the lattice model, we have formulated a dynamic DFT which describes the results of corresponding KMC simulations
very well. In the version used here, growth depends only on the microscopic rate of rods standing up or lying down, i.e. the
rotational mobility.
This independence of translational diffusion through hopping on the substrate has been confirmed by KMC in the case of
neutral substrates, whereas for attractive substrates, growth in KMC depends on hopping diffusion---the DFT results
describe the case of large hopping rates. This particular influence of the substrate is interesting and should be checked 
in further studies, both experimentally and theoretically (for models beyond hard rods). It also points to necessary improvements
in the dynamic DFT treatment. Instead of considering only the rates of change between the averaged densities for lying or standing rods,
the explicit space-- and time--dependence of pair correlation functions in the layer should be calculated, and the averaged densities 
reconstructed from those. 
It is likely that the time--dependent correlation functions are affected by hopping diffusion. 
The inclusion of spatial dependence will also allow for a connection with both the standard dynamic DFT equation for isotropic particles 
in the continuum \cite{Mar99,Arc04} and extensions derived for anisotropic particles \cite{Dho03,Dho05,Rex07,Loe10,Witt11}.  
It would be desirable for the continuum modeling to use FMT functionals for hard spherocylinders having been been developed 
over the past years \cite{Han09,Witt15}.
}  

{
The comparison of the off--lattice, continuum models with hard spherocylinders shows that qualitative agreement in the time--evolution
of nematic order is obtained. This is true once the relevant characteristic times for diffusion, relaxation of nematic order and deposition are matched.
The evolution of the total monolayer density is mainly determined by the equation of state, which differs between lattice and
continuum.   
}

{
For this simple system, we have reached a good methodological control with the lattice and continuum treatments, allowing for
the study of equilibrium, dynamic effects and their interplay. In our opinion, this should be continued in the study of more
complicated and detailed models, and also for studying multilayer growth. With anisotropic rods, the rules for allowed processes
in a KMC lattice formulation are not clear from the beginning, hence continuum simulations are needed to ``gauge'' the dynamic lattice models.
Studies in this direction are in progress. 
}

\section{Acknowledgments}

This work is supported within the DFG/FNR INTERproject ``Thin Film Growth'' by the Deutsche Forschungsgemeinschaft (DFG), Project No. OE 285/3-1
and {SCHR 700/24-1}, by the Fonds National de la recherche (FNR) Luxembourg, and by the Landesgraduiertenf\"orderung Baden--W\"urttemberg. Data from computer simulations presented in this paper were produced using the HPC facilities of the University of Luxembourg~\cite{Var14}.

\begin{appendix}

\section{KMC implementation}
 \label{app:kmc}
 
\begin{figure}
\epsfig{file=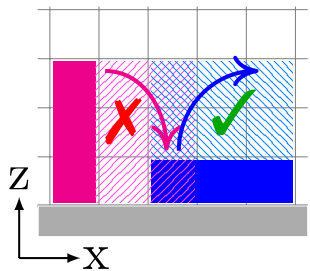, width=4cm}
\caption{(color online) Illustration (out--of--plane) of two neighboring rods and the space they need for rotations about their ends. The left rod is blocked by the right rod, while the right rod is free to rotate: their blockage is not mutual.}
 \label{fig:kmc_alg}
\end{figure}

We implemented a rejection--free KMC algorithm developed in the spirit of Bortz et al.~\cite{BKL} for highly anisotropic hard particles.  
We use a detection system for tracking all allowed/forbidden events in current configurations that is (1) on--the--fly during simulation and (2) 
localized around the change in configuration during each MC step. We restrict our discussion in the following to purely hard--core interactions between 
particles, although these considerations may be extended to finite--ranged interaction potentials. Viewing the kinetics from the point--of--view of a 
particle, a neighbor may exclude one of the particle's elementary moves (translations or rotations) if the neighbor gets close enough. A similar statement 
holds from the point--of--view of the neighbor. According to the KMC method, any of their excluded moves are removed from the current list of possible events. The opposite may also be true---moves may suddenly become possible if the particles have moved apart. These moves must be added to the current list of events. The act of forbidding or allowing the moves of a neighbor is not commutative for anisotropic particles, in general. Fig.~\ref{fig:kmc_alg} shows this situation for rotations of hard rods about their endpoints. 
This non--mutual relationship between neighbors
makes neighbor--lists unsuited for implementation. We outline a characteristically different method using what we denote as {`inverted list indices'} below.

\begin{figure}
\includegraphics[width=7cm]{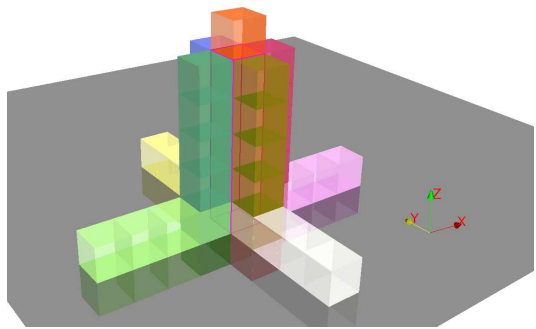}
 \caption{(color online) The {inverted list indices} in our lattice model for the moves of a standing rod with $L=5$: represented are their fields over discrete space, each colored differently. The moves correspond to rotations downwards about the rod--end on the substrate (gray), as well as translational hops to nearby planar sites. Spurious pattern--checking above and beneath the plane is done since the implementation was originally developed for multilayer growth.}
  \label{fig:halo}
\end{figure}

We first take advantage of one feature unique to lattice systems: sites can be tabulated. We implement a field over the lattice that represents the state of occupancy at each site. Given this setup, each particle need only know the local neighborhood pattern of occupancy around it. To clarify, a move by a particle is only possible if a minimal finite volume around it is unblocked by other particles. In addition, if simulations are not restricted to a plane, for example, a move may also require particular sites around it to be occupied (such as in multilayer growth with rods, where a particle may only rotate and translate with occupied sites beneath it as to exclude forming overhangs). Hence, each move by a particle needs this particular pattern around the particle in order to be considered \emph{allowed}; else, the move is \emph{forbidden}. The abstract object representing the tracker for this neighborhood pattern is the {inverted list index}---it acts as a local field over the lattice, moving with the particle and switching with the particle's orientation, accordingly (see Fig.~\ref{fig:halo}). Any change occurring on the lattice is evaluated by the affected {inverted list indices}, and if one changes its state (\emph{allowed} to \emph{forbidden}, or vice versa) the {inverted list index} removes its `index' from the list of allowed events. We note that we adapt the nomenclature of inverted lists from computer science; for the case of $(1\times1)$ particles, Ref.~\cite{Saum09} illuminates the situation: A list of events $\lbrace e_k\rbrace$ is stored, and a particular event $e_k$ is executed at spatial index $(i,j)$ on the lattice (position of a particle). An inverted list $\lbrace e_{(i,j)}\rbrace$ should allow one to quickly access index in memory of the event occurring at position $(i,j)$. This is useful when doing updates locally around the place of each event. Our {inverted list indices} differ in that they exhibit spatial extent and are rather more sophisticated; they perform their updates themselves, i.e. they may add or remove their own indices from the events list. They are merely called to re--evaluate their state if an occupancy has changed within their local field.

\section{Derivation of $\DtwoDlatt$ in the dilute limit}
\label{app:derivation_D2Dlatt}

We begin by writing down the master equation describing the change of a population density $\rho_i(t)$ ($i=1..3$) over time within the dynamics of an ideal lattice gas of tumbling rods (pure hopping does not change the number of rods in any orientation):

\bea
\label{eq:master}
\frac{d \rho_i(t)}{d t} = \sum\limits_{j\neq i}{- \rho_i(t)T(i \rightarrow j) + \rho_j(t)T(j \rightarrow i)}\;,
\eea
where $T(i\rightarrow j)$ is the transition rate for a rod to go from orientation $i$ to $j$. We are interested in steady--state ensemble 
properties---hence, we enforce that for all $i$ the left hand side of Eq.~\eqref{eq:master} is zero and the populations reach a stationary 
state $\lbrace \rho_1,\rho_2,\rho_3\rbrace$. This leaves us with the following condition for the transition rates:\footnote{which can be interpreted as global balance in a Markov--chain Monte Carlo algorithm} 
\bea
\sum\limits_{j\neq i}{\rho_i T(i \rightarrow j) }= \sum\limits_{j\neq i}{ \rho_j T(j \rightarrow i)}\;.
\eea
The transition rates are simply $T(1\rightarrow 2) = T(2 \rightarrow 1) = 2\kxy$, $T(1\rightarrow 3)=T(2\rightarrow 3) = 2\ku$, and $T(3\rightarrow 2) = T(3\rightarrow 1) = 2\kd$, where the factor 2 arises because the rods can rotate into each orientation with positive and negative rotation directions. One can easily show that the only linearly independent equation that remains is the following:
\bea
\label{eq:balance}
(\rho_1 + \rho_2)\ku = 2 \rho_3 \kd 
\eea
Notice that any dependency on the in--plane rotations with rate $\kxy$ drops out of the equations. 
{Defining $\rho_{12}:=\rho_1+\rho_2$, we obtain} $\rho_{12} = \rho_3 \frac{2\kd}{\ku}$. 
Since {the total density is preserved, $\rho= \rho_{12} + \rho_3={\rm const.} $, we find for the stationary state:
\bea
\label{eq:rhos1}
\frac{\rho_3}{\rho} = \frac{1}{1+2\frac{\kd}{\ku}}\\
\frac{\rho_{12}}{\rho} = 1-\frac{1}{{1+2\frac{\kd}{\ku}}}\;.
\label{eq:rhos2}
\eea
}
These equations will be useful in steps that follow.

Returning to expressing the diffusion constant on an infinite 2D lattice, we first consider the motion of the rods of length $L\unit$ only 
along one axis. The first contribution to diffusion comes from {a tumbling move in an average time  $1/\ktum$ (where $\ktum$ is $\ku$ or $\kd$)
which displaces the center--of--mass of the rods by $(L-1)/2$ in units of $\unit$. According to Fick's law for 1D diffusion with diffusion 
constant D, $\langle |\Delta x|^2\rangle = 2 D t$}  we obtain the 1D contribution to translational diffusion from tumbling as {$D = \frac{1}{8}\ktum (L-1)^2\, \unit^2$}.

We next consider specifically the tumbles contributed by upright rods ($i=3$): The mean waiting time for the propagation to this 
mean--squared--displacement is $1/{\kd}$, as before. Therefore, this part of the 1D diffusion, is 
{$\frac{\rho_3}{\rho} \frac{(L-1)^2}{8} \kd  \,\unit^2$, where we included the probability $\rho_3/\rho$ for a rod being upright}. 
$\rho_{12}$ population (on average) contributes to the diffusion in 1D (that half aligned along the corresponding 1D axis). 

The second contribution to translational diffusion along a line is simply the rate $\frac{1}{2}\khopzero$ since, as before, the mean waiting time for the propagation of $1\unit^2$ is $1/{\khopzero}$ (and the same rate is assigned for all populations $\rho_i$).

In summary, the 1D--translational diffusion coefficient on a lattice in units of space $\unit$ is:
\bea
\DoneDlatt/\unit{}^2 = \frac{1}{2} \left(\left( \rho_3 \kd + \frac{1}{2}\rho_{12}\ku \right) \frac{(L-1)^2}{4\rho} + \khopzero\right) \;, 
\eea
where in 1D, the stationary {densities} are $\frac{1}{2}\rho_{12} = \rho_1 = \rho_3$.
{Inserting (\ref{eq:rhos1}) and (\ref{eq:rhos2}) for the density ratios and observing that diffusion} in 2D is simply twice the 
diffusion in 1D,\footnote{The MC moves occur with the same rates in both axes $\hat{x}, \hat{y}$; 
we effectively add the independent Poisson processes together to a process twice as fast in propagating the mean--squared displacement.} we obtain: 
\bea
\DtwoDlatt =  \left( \frac{1}{1+2\frac{\kd}{\ku}}\kd + \frac{1}{2}\left(1-\frac{1}{1+2\frac{\kd}{\ku}}\right) \ku \right)\frac{(L-1)^2}{4} + \khopzero \;,
\eea
 which can be rearranged straightforwardly to the form of Eq.~\eqref{eq:D_latt}.

{\em Verification:}
We verify the form of Eq.~\eqref{eq:D_latt} via KMC simulations of a ideal gas (no interaction energy) of hard rods on an infinite 2D lattice doing hopping and tumbling moves with input parameters $L=9$ and various relations of rates $\lbrace \khopzero, \kd, \ku \rbrace$. The translational diffusion is  measured by fitting a line through the ensemble--averaged mean--squared--displacement over simulated time, given $2\cdot10^4$ rods and some $2^{23}$ MC steps (depending on the relative rates) \emph{after} a certain equilibration time during which each $\rho_i$ reaches a stationary average value. The fit is weighted with the error bars of the data, the error--of--the--mean (ensemble average). The fitted slope corresponds to 4 times the translational diffusion constant, in accordance with Fick's law in 2D. 
A series of such fitted slopes are measured over a few independent trials
{and the agreement with Eq.~\eqref{eq:D_latt} is excellent.} The averaged fitted diffusion rate matches that of Eq.~\eqref{eq:D_latt} to within error bars. The 1D case was verified, as well, where the fitted slope corresponds to twice the diffusion constant.


\end{appendix}

\end{document}